\documentclass[fleqn,usenatbib]{mnras}

\usepackage{mathptmx}
\usepackage{float}

\usepackage[T1]{fontenc}
\usepackage{ae,aecompl}

\usepackage{graphicx}	% Including figure files
\usepackage{amsmath}	% Advanced maths commands
\usepackage{amssymb}	% Extra maths symbols
\usepackage[normalem]{ulem}
\usepackage{soul}
\usepackage{multirow}
\usepackage{mathtools}
\usepackage{lmodern}
\usepackage{xspace} % this decides if space is needed or not

\newcommand{\field}{{\hat \phi}}
\newcommand{\de}[1]{\partial_{#1 }}

\newcommand{\rhoc}{\rho_{c}}
\newcommand{\Rc}{R_{c}}
\newcommand{\Mc}{M_{c}}

\newcommand{\dimR}{\mathrm{Kpc}/\mathrm{h}}
\newcommand{\dimM}{\mathrm{M}_{\odot} / \mathrm{h}}

\newcommand{\PG}{{\small P-GADGET}3\xspace}
\newcommand{\AG}{{\small AX-GADGET}\xspace}

\newcommand{\SUBFIND}{{\small SUBFIND}\xspace}
\newcommand{\AC}{{\small axionCAMB}\xspace}
\newcommand{\citenp}[1]{\citeauthor{#1} \citeyear{#1}}

\newcommand{\main}{halo\xspace}
\newcommand{\sats}{satellites\xspace}

\renewcommand{\vec}[1]{\mathbf{#1}}

\title[Milky-way like system in FDM]{Fuzzy Aquarius: evolution of a Milky-way like system in the Fuzzy Dark Matter scenario}

\author[M. Nori et al.]{
Matteo Nori,$^{1,2}$\thanks{E-mail: matteo.nori@nyu.edu}, Andrea Macciò$^{1,2,3}$ and Marco Baldi$^{4,5,6}$
\\
$^{1}$New York University Abu Dhabi, PO Box 129188 Saadiyat Island, Abu Dhabi, United Arab Emirates\\
$^{2}$Center for Astro, Particle and Planetary Physics (CAP3), New York University Abu Dhabi\\
$^{3}$Max Planck Institut f\"{u}r Astronomie, K\"{o}nigstuhl 17, D-69117 Heidelberg, Germany\\
$^{4}$Dipartimento di Fisica e Astronomia, Alma Mater Studiorum - University of Bologna, Via Piero Gobetti 93/2, 40129 Bologna BO, Italy\\
$^{5}$INAF - Osservatorio Astronomico di Bologna, Via Piero Gobetti 93/3, 40129 Bologna BO, Italy\\
$^{6}$INFN - Istituto Nazionale di Fisica Nucleare, Sezione di Bologna, Viale Berti Pichat 6/2, 40127 Bologna BO, Italy\\
}

\date{Accepted XXX. Received YYY; in original form ZZZ}

\pubyear{2020}
\begin{document}
\label{firstpage}
\pagerange{\pageref{firstpage}--\pageref{lastpage}}
\maketitle

\begin{abstract}

We present the first high-resolution zoom-in simulation of a Milky-way-like halo extracted from the Aquarius Project in the Fuzzy Dark Matter (FDM) framework. We use the N-body code \AG, based on a particle oriented solution of the Schr\"{o}dinger-Poisson equations, able 
to detail the complexity of structure formation while keeping track of the quantum effects in FDM.
The \main shows a cored density profile, with a core size of several kpc for a FDM mass of $m_{\chi}=2.5h \times 10^{-22}\ {\rm eV}/c^2$. A flattening is observed also in the velocity profile, representing a distinct feature of FDM dynamics.
We provide a quantitative analysis of the impact of \textit{fuzziness} on \sats in terms of abundance, mass, distance and velocity distribution functions and their evolution with redshift.
Very interestingly, we show that all collapsed structures, despite showing a flat density profile at $z=0$, do not reach the \textit{solitonic} ground-state at the time of formation: on the contrary, they asymptotically converge to it on a timescale that depends on their mass and formation history.
This implies that current limits on FDM mass -- obtained by applying simple scaling relations to observed galaxies -- should be taken with extreme care, since single objects can significantly deviate from the expected asymptotic behaviour during their evolution.

\end{abstract}


\section{Introduction}
\label{sec:intro}

Among the many unanswered questions about the fundamental processes governing the evolution of our Universe, the ones regarding the elusive nature of dark matter are some of the most puzzling. Its extremely weak interaction (or a lack thereof) with standard matter and its low intrinsic velocity are the two main properties of the standard Cold Dark Matter (CDM) cosmological model, both necessary to avoid unobserved electromagnetic dark matter signals and to be consistent with the formation of large scale structures. These properties, however, can be associated with several physical entities or reproduced by several processes. 

In fact, many possible sources have been historically investigated --~with different success~-- as possible candidates for dark matter \citep[see e.g.][]{Bertone_Hooper_Silk_2005} , such as astrophysical objects \citep[e.g. like Massive Compact Halo Objects (MACHOs)][]{Alcock93}, physical effective mechanisms \citep[e.g. like Modified Newtonian dynamics (MOND)][]{Milgrom83}, as well as new fundamental particles beyond the standard model of particle physics, like Weakly Interactive Massive Particles \citep[WIMPs][]{Jungman95}. 

The WIMP model, in particular, was regarded as the most promising model of dark matter for many years, due to the following peculiar coincidence: assuming the interaction between dark and standard matter to be weak enough to avoid electromagnetic detection but strong enough to ensure thermal equilibrium in the early universe before the eventual decoupling of the dark component, the mass of such dark matter particle would fall in the $100-1000$~GeV$/c^2$ range, which is coincidentally similar to the range associated to weak force \citep[see e.g.][]{Kamionkowski97}.

With the advent of new cosmological surveys of the Cosmic Microwave Background like WMAP \citep{wmap9} and PLANCK \citep{Planck15}, the WIMP particle-based model was still considered as one of the front-runner models for dark matter. Nevertheless, the lack of a detection in this mass range after many years of continuous search by dedicated particle accelerator experiments like LHC \citep[see e.g.][]{Fermi17annih, Danninger17, Buonaura18} has been eroding the consensus surrounding WIMPs, in favor of other particles in mass ranges previously unexplored. 

\bigskip

Among these, an interesting dark matter candidate could be linked to the axion particle \citep[see again][]{Kamionkowski97}, arising from the CP-symmetry breaking in Quantum-Chromodynamics (QCD) \citep{PecceiQuinn77a,PecceiQuinn77b}.
From the QCD axion -- which is tightly related to the strong CP problem -- a more general concept of a pseudo-scalar bosonic particle can be derived to encompass a much broader class of axion-like particles (ALPs). Within this category spanning over an astonishingly wide range of masses of the order $10^{-24} - 10^{0}$ eV$/c^2$, there are particles that can be regarded as well-motivated dark matter candidates well below the WIMP mass scale \citep[see][for a recent and comprehensive review on the subject]{Ferreira21review}.

The general ALPs dynamics is characterised by a wave-like quantum self-interaction that is mainly set by its mass. The latter is in fact related both to the timing of the dynamical regimes ALPs exhibit with respect to the background cosmic dynamics as well as to the the strength of the self-interaction and its consequences on structure formation. In particular, the ALPs mass sets the cosmological epoch at which the associated dark matter component exits from the oscillatory regime --~which is a peculiar feature of the axion potential~-- and begins to cluster, thus affecting in different ways the evolution of large scale structures when ALPs role as a dark matter candidate is considered \citep[see e.g.][]{Sikivie08}.

A crucial distinction concerning the relative timing of the end of the oscillatory regime with respect to the time of matter-radiation equality can be then drawn for ALPs with mass above or below $10^{-10} \rm{eV}/c^2$. Above this value, dark matter begins to cluster before equality thus effectively segregating a large fraction of the total dark matter content in gravitationally bound \textit{axion miniclusters}
by the time of baryon decoupling from radiation \citep[see e.g. the early works of][]{Kolb93,Kolb94}.
On the contrary, for lower values of the mass the density distribution of the ALP field at matter-radiation equality can be essentially described by adding a small-scale correction to the usual CDM density distribution related to its self-interaction \citep[][]{Hu00}.

The clustering process differs in the ALP dark matter and CDM models, as the typical wave-like dynamics of the ALP acts as an effective net repulsive force below a certain scale, thus admitting a non-degerate self-gravitating stable solution --~called \textit{soliton}~--, whose properties scale with the ALP mass \citep[see e.g.][]{Marsh16review}.
In recent years, a wide range of experiments have been designed to detect QCD axions or ALPs and to investigate their possible link to dark matter \citep[see e.g.][for a recent overview]{Banerjee19}: these include e.g. resonant cavity experiments at various frequencies (ORGAN \citenp{ORGAN}, ADMX \citenp{ADMX20}) dielectric haloscopes \citep[MADMAX][]{MADMAX17}, detection-induced magnetic flux oscillations \citep[ABRACADABRA][]{ABRACADABRA19} and NMR-based techniques (CASPEr \citenp{CASPER} and ARIADNE \citenp{ARIADNE}).

\bigskip

In this work, we focus on the lower end of the ALPs mass spectrum --~in the range of $10^{-24}-10^{-19} {\rm eV}/c^2$~-- whose associated dark matter models are often referred to as Fuzzy Dark Matter (FDM). Masses in this range are often conveniently expressed in units of $m_{22}$, where $m_{22}\equiv10^{-22} {\rm eV}/c^2$. As previously mentioned, the wave-like interaction of FDM acts as a net repulsive force and modifies the standard matter power spectrum of CDM at matter-radiation equality, effectively smoothing out density perturbations at small scales and thus leading to fewer collapsed structures at lower redshifts \citep{Hu00}. Moreover, the particle mass is so light that the associated de Broglie wavelength and --~as a direct consequence~-- the self-gravitating objects that can be formed are comparable with the galactic scales \citep[see again][]{Hu00}. These features are of particular cosmological interest, since FDM could be simultaneously involved in the development of a constant-density core in the innermost region of dark matter profiles --~originally related to the cusp-core differentiation of \citet{Oh11}, that can be solved only in some mass ranges by the presence of baryons  \citep[see e.g.][]{Tollet15}~-- and the suppression of the typical CDM overabundance of galactic satellites \citep{Klypin_etal_1999}.

Numerical simulations of structure formation within FDM models have been initially performed by means of highly numerically intensive Adaptive Mesh Refinement (AMR) algorithms able to solve the Schr\"odinger-Poisson equations over a grid \citep[see e.g.][]{GAMER,GAMER2,Mocz17}, leading to impressive and very detailed results on the properties of individual FDM collapsed objects \citep[see e.g.][]{Woo09,Schive14,Veltmaat18}. However, the computational cost of such approach hindered the possibility to extend the investigation of late time structure formation to large cosmological volumes. To address this issue, N-Body codes were developed, initially only including the (linear) suppression in the initial conditions but neglecting the integrated effect of the FDM interaction during the subsequent dynamical evolution \citep[see e.g.][]{Schive16,Irsic17,Armengaud17} --~i.e. basically treating FDM as standard dark matter with a suppressed primordial power spectrum, similarly to what is routinely done in Warm Dark Matter simulations \citep[][]{Bode00,Schneider_etal_2012}~--.

\bigskip

In this manuscript, we present a high-resolution Milky-way-like dark matter system simulated in the Fuzzy Dark Matter scenario; in particular, the system consists in a zoom-in simulations taken from the Aquarius project \citep{Aquarius} (halo A - resolution level 3). The Aquarius project has been a cornerstone in understanding the relationship between halo and subhaloes properties, and it has been used as a benchmark for investigating other dark matter models beyond CDM in the past \citep[as e.g. Warm Dark Matter][]{Lovell13}. The main goal of this work is the analysis of the system properties and FDM specific observables and their evolution in time as resulting from the complex non-linear structure formation process in a FDM cosmology. 

The zoom-in approach consists in a rationalised distribution of resolution elements within the simulation box, which allows to detail a region of interest --~normally, a collapsed structure~-- with high-resolution while efficiently keeping track of its environment \citep[see e.g.][]{Navarro94,Katz94}. In this sense, zoom-in simulations represent an intermediate step bridging single-object simulations and bigger fixed-resolution cosmological simulations. By adopting this method, it is then possible to retain information about the cosmological context while reaching a higher resolution than a homogeneously represented N-body simulation.

\bigskip

The paper is organised as follows: in Sec.~\ref{sec:theory} we present the theoretical background related to FDM models; in Sec.~\ref{sec:NM} we detail the numerical aspects of this work, related to the simulation and the data production; in Sec.~\ref{sec:results} we then present the main results of this work, regarding the \main and \sats; finally, in Sec.~\ref{sec:conclusions} we summarise our findings, discuss them and draw our conclusions.

\section{Theory}
\label{sec:theory}

In this Section, we collect the fundamental equations that govern FDM dynamics. We here detail also the specific properties and scaling relations that are relevant to this work and that characterise FDM collapsed structures.

\subsection{Fuzzy Dark Matter models}
\label{sec:fdm_th}

In FDM models, the mass of the dark matter particle is so tiny that the associated De-Broglie wavelength is of astrophysical scales, requiring the dynamical treatment of dark matter to take into account quantum interactions. FDM is thus usually described through a quantum bosonic field $\field$, under the assumption of condensation \citep{Hu00,Hui16}.

The Gross-Pitaevskii-Poisson equation describing the evolution of a massive bosonic field $\field$ reads \citep[][]{Gross61,Pitaevskii61}
\begin{equation}
\label{eq:GPP}
i \frac{\hslash}{m_\chi} \ \de{t} \field = - \frac {\hslash^2} {m_{\chi}^2} \nabla^2 \field + \Phi \field % + \lambda \left| \field \right|^2 \field
\end{equation}
where $m_{\chi}$ represents the typical mass of the FDM particle and $\Phi$ is the Newtonian gravitational potential, % $\lambda$ is the self-interaction coupling constant
that satisfies the standard Poisson equation 
\begin{equation}
\label{eq:poisson}
\nabla^2 \Phi = 4 \pi G \rho_b \ \delta / a
\end{equation}
where $\delta=(\rho - \rho_b)/\rho_b$ is the comoving density contrast with respect to the comoving background density $\rho_b$ \citep{Peebles80}. Together, Eq.~\ref{eq:GPP} and Eq.~\ref{eq:poisson} form the so-called Schr\"{o}dinger-Poisson (SP) system.

The system can be recast from a \textit{field} description into a mathematically equivalent \textit{fluid} one --~associating the field amplitude and phase with fluid density $\rho$ and a fluid velocity $\vec v$, respectively~-- with the use of the Madelung transformation \citep[][]{Madelung27}
\begin{gather}
\rho = \left| \field \right|^2 \\
\vec v = \frac {\hslash} {m_\chi} \Im{ \frac {\vec \nabla \field} {\field} }.
\end{gather}

In the frame of an expanding universe --~with $a$ and $H=\dot a / a$ being the usual cosmological scale factor and Hubble function, respectively~--, we refer to $\vec x$ as the comoving distance and to the velocity $\vec u$ as the comoving equivalent of $\vec v$. The real and imaginary parts of Eq.~\ref{eq:GPP} then translate into a continuity equation
\begin{equation}
\label{eq:continuity}
\dot \rho + 3 H \rho + \vec \nabla \cdot \left( \rho \vec u \right) =0
\end{equation}
and a modified Euler equation
\begin{equation}
\label{eq:NS}
\dot {\vec u} + 2H \vec u + \left( \vec u \cdot \vec \nabla \right) \vec u =  - \frac {\vec \nabla \Phi} {a^2} + \frac {\vec \nabla Q} {a^4}
\end{equation}
where an additional potential $Q$ --~accounting for the wave-like behaviour of the field~-- appears alongside the usual gravitational potential $\Phi$.  

The so-called Quantum Potential $Q$ (QP hereafter) has the form
\begin{equation}
\label{eq:QP}
Q = \frac {\hslash^2}{2m_{\chi}^2} \frac{\nabla^2 \sqrt{\rho}}{\sqrt{\rho}}  = \frac {\hslash^2}{2m_{\chi}^2} \left( \frac {\nabla^2 \rho} {2 \rho} - \frac {| \vec \nabla \rho|^2}{4 \rho^2} \right)
\end{equation}
and accounts for the purely quantum behaviour of the field \citep{Bohm52}. It is interesting to remark that, from a theoretical point of view, the quantum nature of dark matter that sources the QP is present in the usual Euler equation used to describe CDM as well: however, it is just safely negligible in the classical limit, as the factor $\hslash^2/m_{\chi}^2$ is extremely small for the typical eV-GeV mass ranges that has been historically considered for the CDM particle \citep[see e.g.][]{Bertone_Hooper_Silk_2005,Feng10}. 

\subsection{Fuzzy Dark Matter: scaling relations}
\label{sec:fdm_sr}

The modified Euler-Poisson (mEP) system composed by Eq.~\ref{eq:continuity}, Eq.~\ref{eq:NS} and Eq.~\ref{eq:poisson} that governs self-gravitating FDM dynamics
\begin{equation}
\label{eq:mEP}
\begin{dcases}
\dot \rho + 3 H \rho + \vec \nabla \cdot \left( \rho \vec u \right) = 0 \\
\dot {\vec u} + 2H \vec u + \left( \vec u \cdot \vec \nabla \right) \vec u =  - \dfrac {\vec \nabla \Phi} {a^2}+ \dfrac {\vec \nabla Q} {a^4} \\
\nabla^2 \Phi = 4 \pi G \rho_b \ \delta / a
\end{dcases}
\end{equation}
is the fluid-equivalent of the SP system. It admits a spherically symmetric time-independent one-parameter family of solutions $\rho_{\rm sol}(r)$ that has no analytical form. The ground-state solution is usually referred to as the \textit{solitonic} core, as its density profile saturates to a constant value in the centre. Despite the lack of an analytical form, it can be well approximated \citep[see e.g.][]{Schive14} by
\begin{equation}
\label{eq:soliton}
\rho_{\rm sol}(r, \rho_c, R_c) = \rhoc \left[ 1 + \alpha \left( \frac{r}{\Rc} \right)^2 \right]^{-8}
\end{equation}
where $\rhoc$ is the core density
and
\begin{equation}
\label{eq:Rc}
\Rc : \rho_{\rm sol}(r=\Rc) = \rhoc / 2
\end{equation}
is the half-density comoving radius, simply referred to as core radius, that is set by construction choosing the constant $\alpha  = \sqrt[8]{2}-1$. 

The mEP system of Eq.~\ref{eq:mEP} is invariant under the coordinate transformation via a generic constant $\lambda$ \citep{Ji94}
\begin{equation}
\begin{split}
\label{eq:transformation}
\{ \vec{x}, t,& \vec{u}, \rho, M, \Phi, E \} \\
&\rightarrow \left\{ \lambda \tilde{\vec{x}}, \lambda^2 \tilde{t}, \lambda^{-1} \tilde{\vec{u}}, \lambda^{-4} \tilde{\rho}, \lambda^{-1} \tilde{M}, \lambda^{-2} \tilde{\Phi} , \lambda^{-3} \tilde{E} \right\}
\end{split}
\end{equation}
where we also included the mass $M$ and the energy $E$ of the system.

It is possible to see that such transformation sets some scaling relations, in particular the core density $\rhoc$, its radius $\Rc$ and its mass $\Mc$ are thus linked through
\begin{equation}
\label{eq:scaling1}
\Rc \propto \left( a \ m_{\chi}^2 \ \rhoc \right)^{-1/4}
\end{equation}
%and --~using Eq.~\ref{eq:mass_soliton}~--
%\begin{equation}
%\label{eq:scaling1bis}
%\Rc \propto \left( a \ m_{\chi}^2 \ \Mc \right)^{-1}
%\end{equation}
thanks to the intrinsically symmetric nature of the system \citep[see e.g.][for a thorough analytical and numerical study]{Chavanis11a,Chavanis11b}. 

In this work, we specifically use the term FDM cores to describe the one-parameter discreet family of spherically symmetric solutions --~each associated with a different energy level~-- that satisfy the scaling relation $\Rc \sim (a \rhoc)^{-1/4}$, with the \textit{solitonic} core representing the densest ground-state solution \citep[see e.g. Appendix B in][]{Hui16}.

In a cosmological context, the linear density perturbation $\delta_k$ in Fourier space satisfies 
\begin{equation}
\label{eq:PERT}
\ddot \delta_k + 2 H \dot \delta_k + \left(  \frac{\hslash^2 k^4}{4 m_{\chi}^2 a^4} - 4 \pi G \rho_b \right) \delta_k = 0
\end{equation}
that directly sets the typical scale 
\begin{equation}
\label{eq:kq}
k_Q (a)= \left( \frac{16\pi G \rho_b m_{\chi}^2}{\hslash^2} \right)^{1/4} a^{1/4} 
\end{equation}
for which the gravitational pull is balanced by the QP repulsion, sometimes referred as \textit{quantum Jeans scale} in analogy with the homonym classical one \citep{Chavanis12}. This scale decreases as the Universe evolves, effectively shifting the scale of balance between quantum repulsion and gravitational attraction towards smaller and smaller values. 
As a result, dark matter structures that are able to collapse are thus characterised by a central core in the density profile, as the quantum interaction is able to sustain it against gravitational collapse even at lower redshifts, while a typical Navarro-Frenk-White (NFW) density profile is recovered at larger distances, where the contribution of quantum interaction is negligible compared to the gravitational one. 

The scaling relation of Eq.~\ref{eq:scaling1} was first explicitly investigated in an astrophysical scenario with dedicated numerical simulations by \citet{Schive14}, where it was confirmed to hold for a sample of haloes at different redshifts in the mass range $10^{9}-10^{11} \dimM$, simulated by directly solving the Schr\"{o}dinger equation on a three-dimensional grid.

Many following works by different groups using a grid-based approach \citep[see e.g.][]{Schwabe16,Du16,Mocz17,May21,Chan21} confirmed the statistical validity of the relation for individual systems. The scaling relation was found to be generally valid also in N-body simulations \citep{Nori20}, although a statistically higher normalisation constant with respect to other works was observed, due to a possible correlation between the normalisation and the dynamical state and/or mass of the specific system.

\section{Numerical Methods}
\label{sec:NM}

In this Section, we introduce and describe the simulations presented in this work and the numerical algorithms used in the simulation process.

\subsection{Zoom-in simulations and IC setup}
\label{sec:sims}

The results discussed in the present work are extracted from two simulations of the same system. In particular, we will systematically compare the main results of one of the haloes of the Aquarius Project \citep[][halo A-3]{Aquarius} --~simulated within the CDM scenario~-- with the ones obtained from its re-simulated counterpart in the FDM context.

The two simulations are \textit{zoom-in} simulations, meaning that their initial conditions are based on a rationalised distribution of resolution elements --~particles, in the case of N-body codes~-- belonging to a collapsed object extracted from a parent low-resolution cosmological simulations. %On a technical level, this is done by tracing back the position of all the particles belonging to a halo in the parent simulations to their original position in the initial conditions to identify the so-called Lagrangian region. This region is then populated with a larger number of less massive particles while the opposite is done outside the region of interest, effectively redistributing the resolution power where it is most desired.

In terms of initial conditions, the distribution of initial density perturbations have been modified accordingly to the FDM case. The typical suppression of the initial power spectrum induced by FDM was taken into account by providing a suitable initial power spectrum for the desired FDM cosmology. To calculate the cosmological initial power-spectrum at the initial redshift, we used \AC \citep{axionCAMB}, assuming the totality of matter to be fuzzy with a boson mass of $m_{22} = 2.5 h$. 

In terms of dynamical evolution, the two simulations effectively share the same background cosmology with the only exception being the dark matter dynamics. The original halo from \citet{Aquarius} was simulated using \PG and its FDM counterpart has been simulated with a modified version of \PG that include FDM phenomenology called \AG \citep{Nori18}. The modified code makes use of the \PG Smoothed Particle Hydrodynamics (SPH) routines to solve FDM dynamics in the Madelung framework, interested readers can find more information on the code in \citet{Nori18} \citep[as well as in][]{Nori19,Nori20}. 

The differences observed in the final properties between the two simulations are then to be considered as consequences of the two main aspects that differentiate FDM from CDM phenomenology. On the one hand, the smoother density distribution of the early FDM universe at small scales, encoded in the simulations as a suppression of the initial FDM power spectrum. On the other hand, the additional QP term in the equations of motion of dark matter.

As a final note, we report the cosmological background parameters used in the FDM simulation as well as in the original CDM one from \citet{Aquarius}: $\Omega_m = 0.25$, $\Omega_{\Lambda} = 0.75$, $H_0 = 73 \ \text{km/s/Mpc}$ together with the initial power-spectrum parameters $n_s = 1$ and $\sigma_8 = 0.9$. 

\subsection{Halo identification, merger tree construction and fragmentation correction}
\label{sec:subfind}

To identify collapsed objects and build the merger tree, we use the same methods (based on the \SUBFIND code) as in \citet{Aquarius}. The \SUBFIND code relies on a Friends-of-friends algorithm \citep[][]{Davis_etal_1985} tailored to identify halo systems, that are then subdivided and disentangled in subhaloes based on binding energy criteria. 

In the following, we will refer to the overall dark matter system centred in the potential minimum as the \main, while we refer to the \main subhaloes as \sats, in clear analogy with the Milky-way and its smaller companions.

The information related to the evolution of structures is encoded by the merger tree, reconstructed from the several snapshots of the simulation taken at different redshifts. Making use of the constant and unique numerical identifier of each particle in the simulation, it is possible to track the evolution of particles ensembles, associating progenitors and descendants (i.e. haloes that share a large number of particles at different redshifts) and identify merger events where multiple haloes coalesce into one. By comparing each snapshot with the previous one, it is also possible to quantify at a given snapshot the share of mass of an object that has been accreted via merger with other objects, quantity that we define as $M_\text{merged}$.

It is important to know that not all the collapsed objects that are present in the FDM simulations can be safely taken in consideration. In fact, a known problem that affects N-body simulations with suppressed initial condition at small scales is the so called \textit{numerical fragmentation} problem \citep[][and references therein]{Wang_White_2007}, that indicates the formation of small collapsed objects that have numerical origin and do not arise from the gravitational evolution of primordial physical over-densities. 

To filter our satellite sample from this spurious contamination, we use the same approach based on cuts in number of particles, mass and shape used in \citet{Nori19} \citep[and based in turn on][]{Wang_White_2007,Lovell13}. In practice, this requires to impose a cut-off in mass, the number of particles and the shape on FDM subhaloes; these cuts are a conservative way to safely regard the remaining sample as physically meaningful. 

The empirical estimate for the mass cut-off scales with the dimensionless power spectrum peak scale $k_{peak}$ and the inter-particle distance $d$ \citep{Lovell13,Wang_White_2007} as
\begin{equation}
\label{eq:MLIM}
M_{CUT} \sim 5\ \rho_b\ d\ /\ k^2_{peak}
\end{equation}
above which it is possible to say that most of the haloes have a physical origin. In the case of our FDM simulation, this value is approximately $M_{CUT} \sim 5 \times 10^8 \dimM$ representing $\sim 10^4$ particles (a factor $10$ higher than the cut-ff in the number of particles suggested by \citet{Lovell13}). As for the shape, the halo particles are traced back to their original position in the initial conditions and the ratio of the minor and major semi-axes --~obtained by computing the inertia tensor of the equivalent triaxial ellipsoid and defined as \textit{sphericity} $s$~-- is used as a constraint. A cut-off is imposed to $s$ in the initial condition, below which the haloes are considered spurious; in this work, we use the value $s_{CUT}=0.16$ of \citet{Lovell13} that we verified being valid also in the FDM case in \citet{Nori19}.

\section{Results}
\label{sec:results}

In this Section, we present the results obtained from the Aquarius FDM zoom-in simulation and we compare it with its CDM counterpart. In the comparison between the two, we highlight the specific effects introduced by the different dark matter behaviour. In particular, we first detail the global properties of \sats, their cumulative distributions in terms of mass, position and velocity in FDM and CDM cosmologies. We then focus on the specific FDM properties of the \main and \sats cores to characterise the evolution of their features.

\subsection{FDM vs CDM comparison of global properties}

\begin{figure*}
%\frame{
\includegraphics[height=0.90\columnwidth,%width=0.49\textwidth,
trim={0.5cm 0.9cm 3.3cm 0cm}, clip] {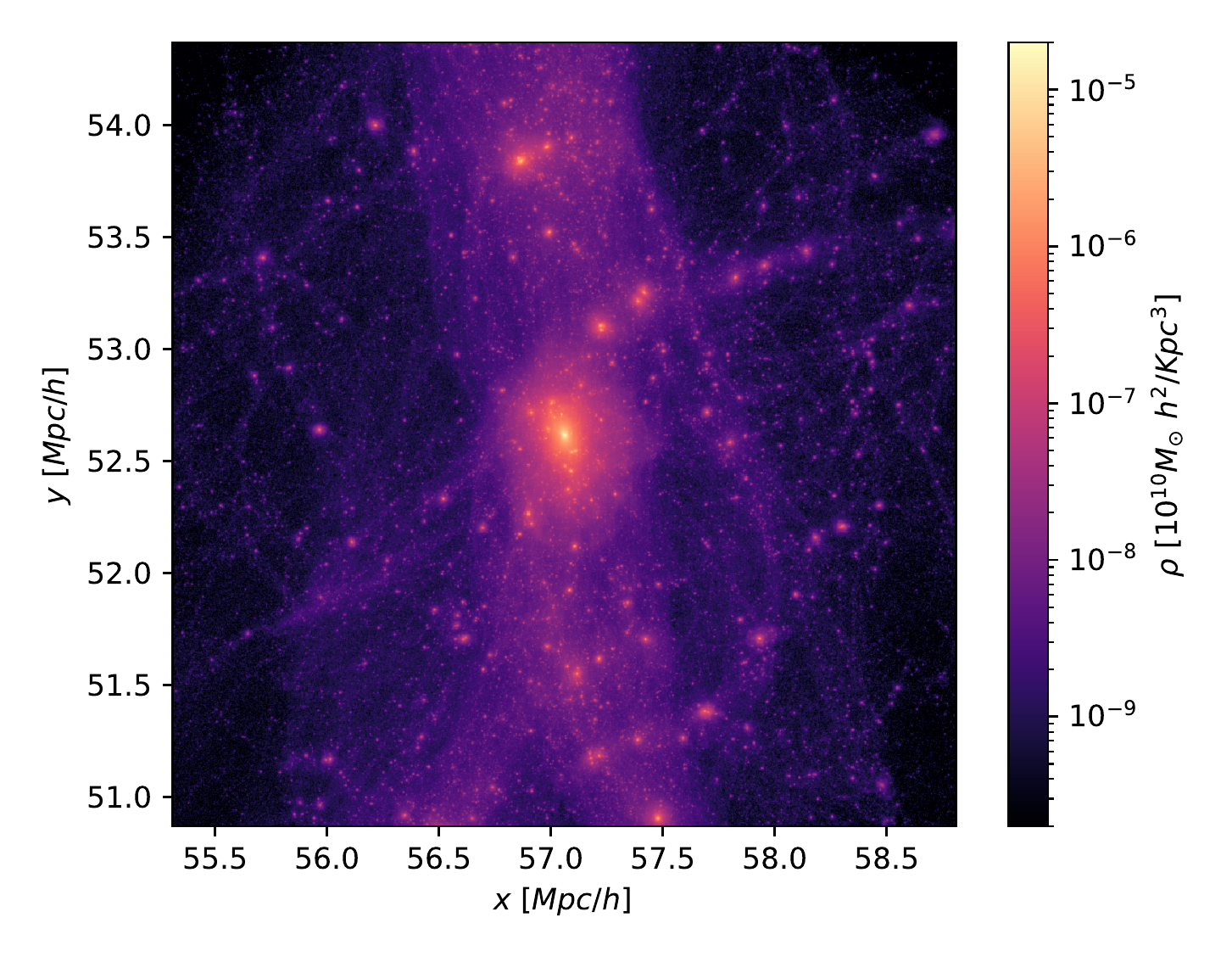}
%}
%\frame{
\includegraphics[height=0.90\columnwidth,%width=0.49\textwidth, 
trim={1.0cm 0.9cm 0.6cm 0cm}, clip] {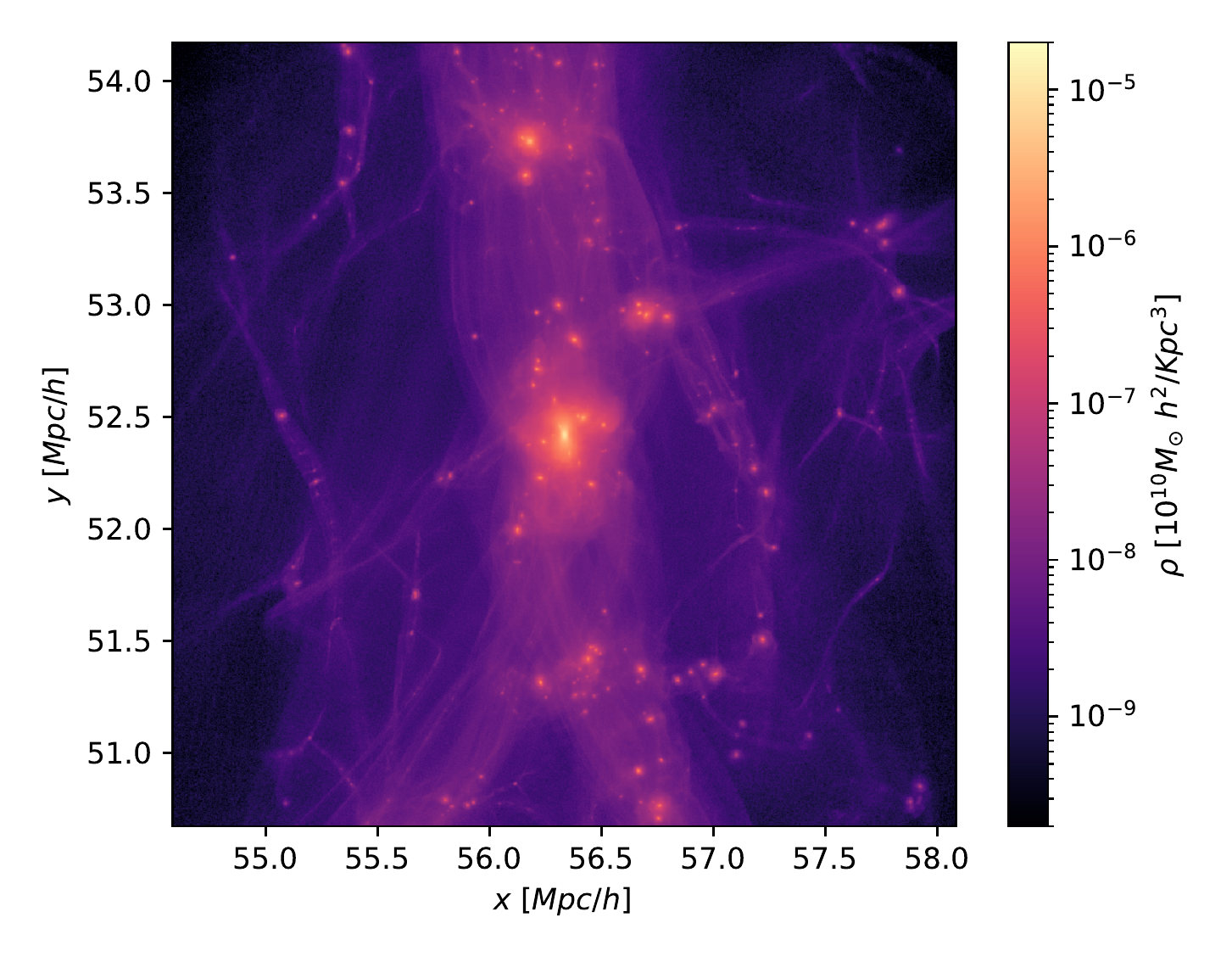}
%}
%\frame{
\includegraphics[height=0.90\columnwidth,%width=0.49\textwidth, 
trim={0.5cm 0.5cm 3.3cm 0.4cm}, clip] {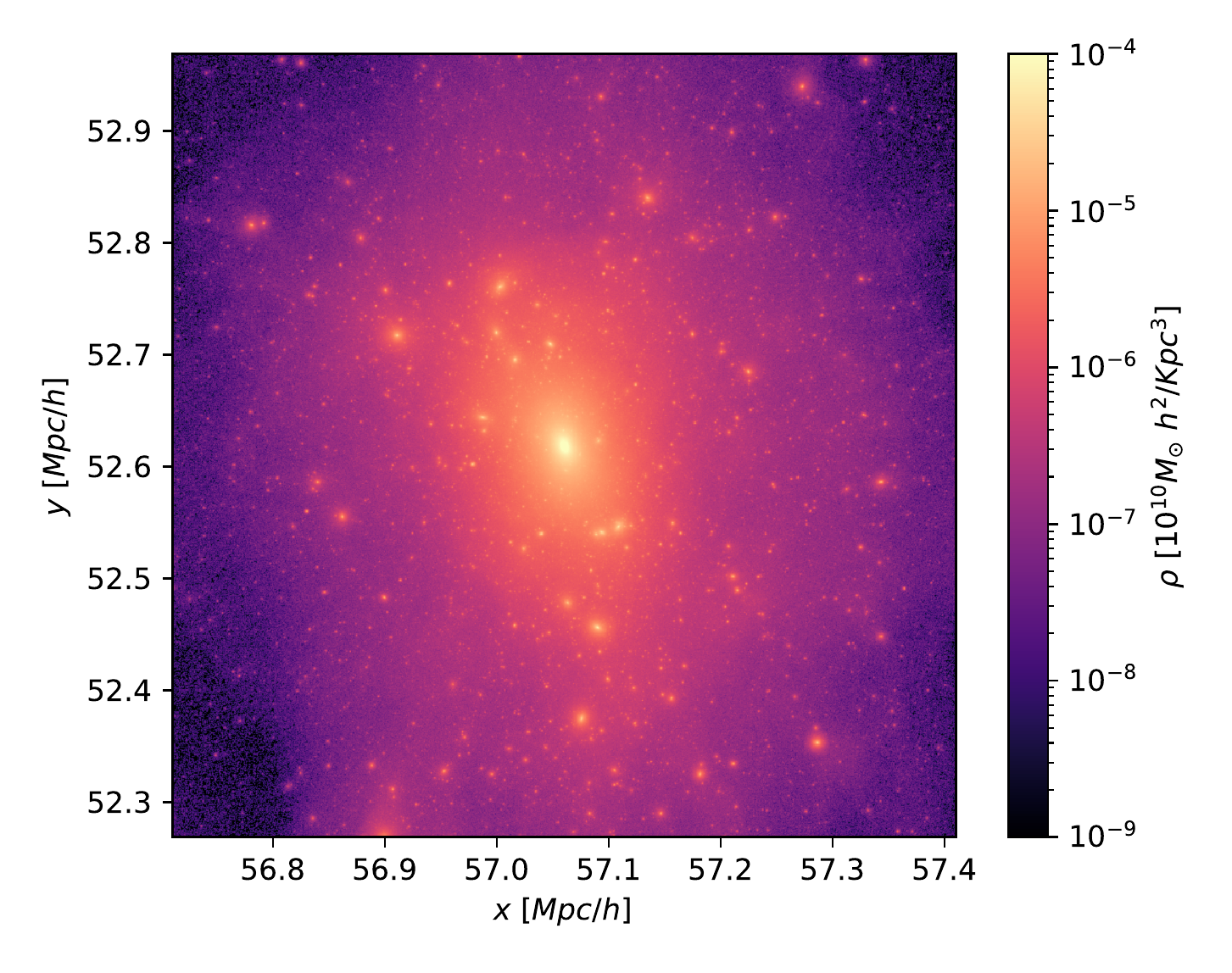}
%}
%\frame{
\includegraphics[height=0.90\columnwidth,%width=0.49\textwidth, 
trim={1.0cm 0.5cm 0.6cm 0.4cm}, clip] {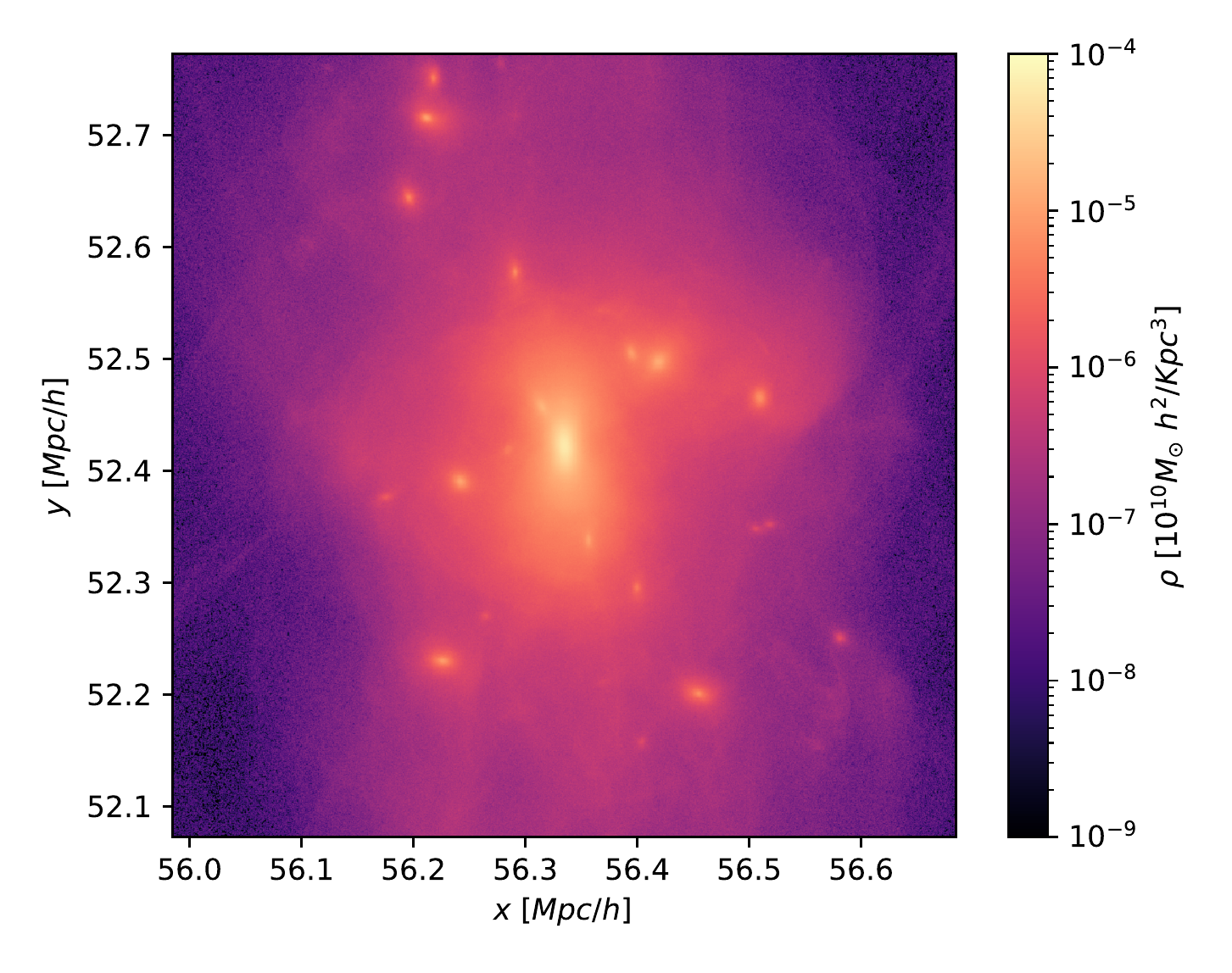}
%}
\caption{Density maps at $z=0$ of the CDM (left panels) and FDM (right panels) of the high-resolution region in a $3.5~\text{Mpc/h}$ side cube (top panels) and a zoom of $0.7~\text{Mpc/h}$ side on the main system (bottom panels).}
\label{fig:maps}
\end{figure*}

\begin{figure*}
\includegraphics[width=0.495\textwidth, trim={2cm 2cm 2cm 2cm}, clip] {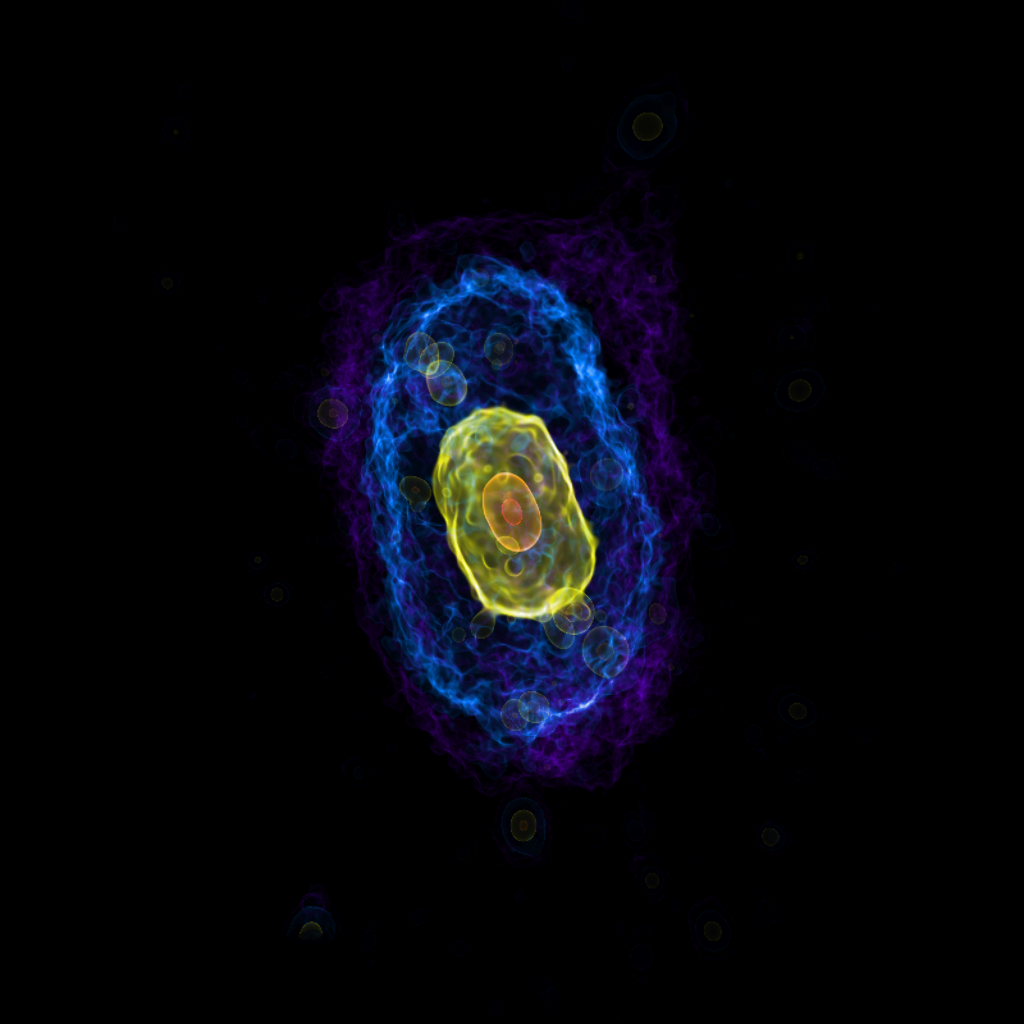}
\includegraphics[width=0.495\textwidth, trim={2cm 2cm 2cm 2cm}, clip] {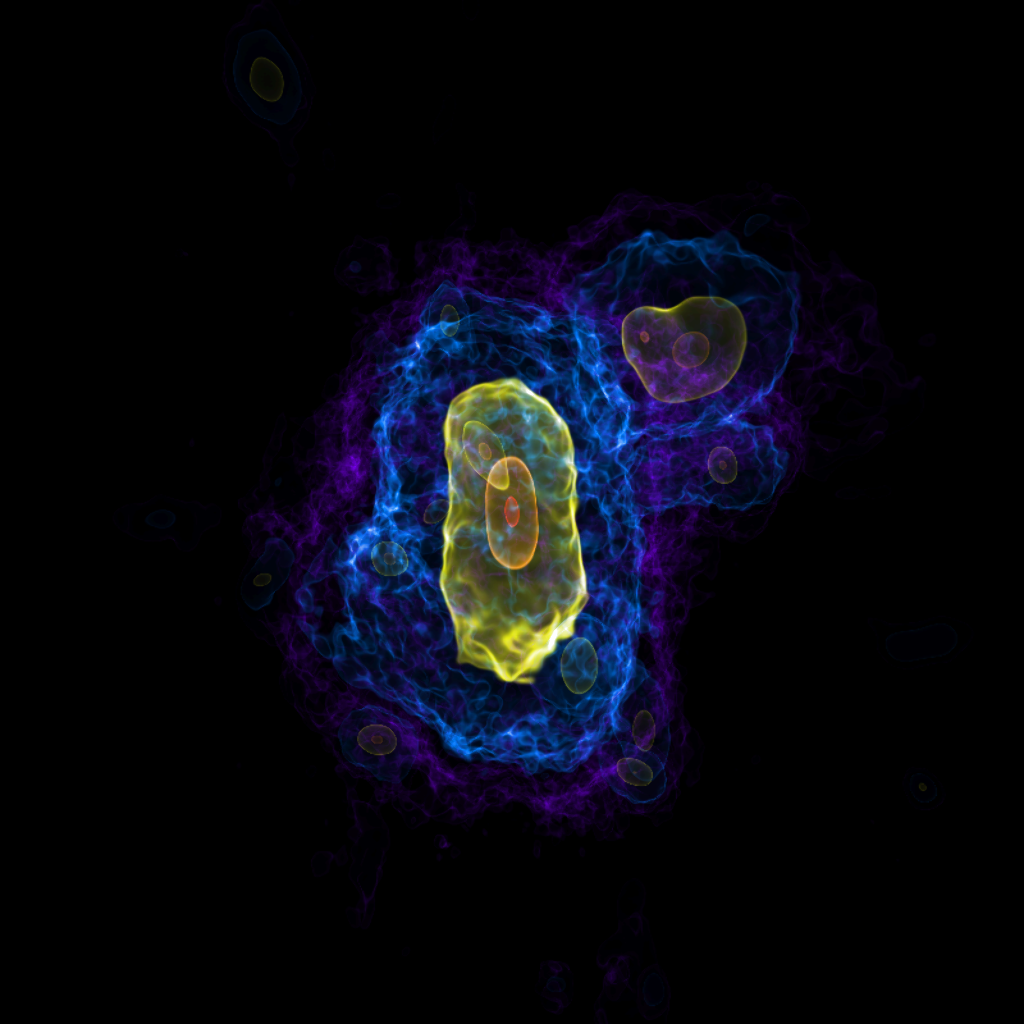}
\caption{Density renderings of the main halo in the CDM (left) and FDM (right) scenario. Colours represent iso-density surfaces at different density values within a box of $1~\text{Mpc/h}$. A movie of the $360^\circ$ rotation around the y-axis can be found in the online supplement material.}
\label{fig:renderings}
\end{figure*}

As a first visual comparison, in Fig.~\ref{fig:maps} we present two density maps at $z=0$ of the CDM (left panels) and FDM (right panels) high-resolution region of $3.5\ \dimR$ per side (top panels) and a zoom of $0.7\ \dimR$  per side on the main system (bottom panels). The difference between the two scenarios in terms of number of structures is striking, especially for the large number of small structures that characterise CDM and are not present in FDM. To highlight this feature in a 3-dimensional space, a density rendering is displayed in Fig.~\ref{fig:renderings}. The images depict the main halo at $z=0$ for the CDM (left) and FDM (right) simulations, within a box of $0.7\ \dimR$ side, using the same iso-density contours level. For the interested readers, a $360^{\circ}$ rotating version of Fig.~\ref{fig:renderings} can be found in the online materials.

The following detailed statistical analyses on the properties of satellites take into account the effect of the numerical fragmentation in the FDM simulation. In fact, not all the FDM systems represent a physical collapsed object as they may be the artificial result of numerical noise in the initial conditions, as discussed in Sec.~\ref{sec:subfind}. %\MN{[Add here number of \sats at $z=0$ just as an example.]} 
Depending on the situation, we will include a comparison with the subsample of CDM \sats that satisfies the same requirement $M_{sat}>M_{CUT}$ (referred as CDM-CUT), to specifically compare FDM and CDM results in the same mass range. 

\subsubsection{Number and mass of \sats}
\label{sec:numbermass}

As visually clear from Fig.~\ref{fig:maps}, the Milky-way like system under consideration presents significant differences in the number of \sats when comparing the FDM and CDM frameworks. 

\begin{figure}
\includegraphics[width=\columnwidth, trim={0.16cm 0.11cm 0.11cm 0.11cm}, clip] {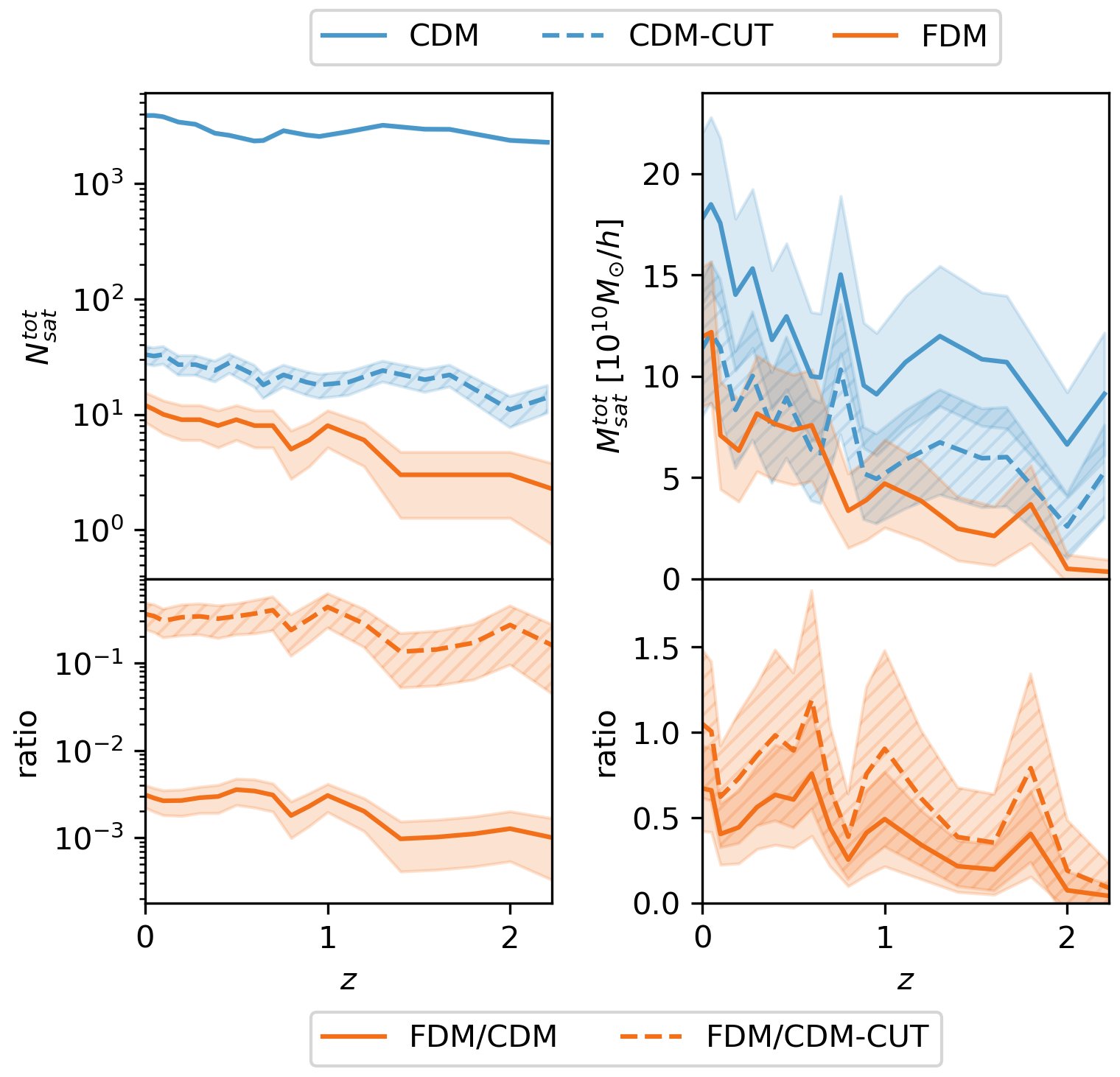}
\caption{Evolution of the total number of \sats (left panels) and their total mass (right panels) in time. The results obtained for the CDM (solid blue), the CDM-CUT (dashed blue) and the FDM (solid orange) samples are displayed in absolute terms in the top panels. Bottom panels show the ratio of the quantities above (solid and dashed lines for the FDM/CDM and FDM/CDM-CUT, respectively). $1 \sigma$ statistical errors are included as shaded regions.}
\label{fig:mbound}
\end{figure}

First of all, let us consider the number of \sats and the mass share they represent with respect to the whole system. These quantities are collected and displayed in Fig.~\ref{fig:mbound} as functions of time. The upper left panel shows the total number of \sats $N^{tot}_{sat}$ found in the FDM and CDM simulations, as well as the subsample CDM-CUT, identified by solid orange, solid blue and dashed blue lines, respectively; the lower left panel represents the ratio between FDM and the CDM and CDM-CUT samples. Using the same colour-coding, the right panels display the total satellite mass contribution $M^{tot}_{sat}$ --~i.e. the sum of the mass of all satellites~--, again presented both in absolute values (upper right panel) and relative terms (lower right panel). Poissonian statistical errors are depicted as shaded areas.

In the FDM scenario there is a great reduction in the total number of \sats with respect to CDM, as expected. In this particular system and for the FDM boson mass $m_{\chi}$ under consideration, the ratio is found to be stable in the $0.1-0.3 \%$ range. When comparing FDM and CDM-CUT \sats belonging to the same mass range, the ratio is two dex larger, approximately $10-30 \%$. These ratios appear to be rather constant throughout the simulation.

The mass share found in \sats is reduced as well; however, in this case a trend is noticeable: the total \sats mass in FDM is approximately $\sim20\%$ the one in CDM around redshift $z\sim2$ but this difference becomes less pronounced in time, with a value of $\sim60\%$ at redshift $z=0$. Restricting the analysis on the same mass range, after an initial suppression, FDM approaches CDM-CUT and the two become statistically consistent after $z\sim0.5$.

The relative number and mass of \sats found in FDM and CDM and their evolution are consistent with two specific physical effects introduced by the FDM cosmology. On the one hand, the typical absence of FDM overdensities at small scales at high redshift directly translates in a lower number of substructures that are eventually able to form; on the other hand, as the quantum Jeans scale becomes smaller and smaller, gravity overtakes the repulsive effect of QP and the most massive \sats~--~that dominate the FDM and CDM-CUT samples~-- are able to non-linearly accrete more and more mass, eventually reaching a similar final mass as their CDM counterparts.

\begin{figure*}
\includegraphics[width=0.49\textwidth, trim={0cm 0cm 0.1cm 0cm}, clip] {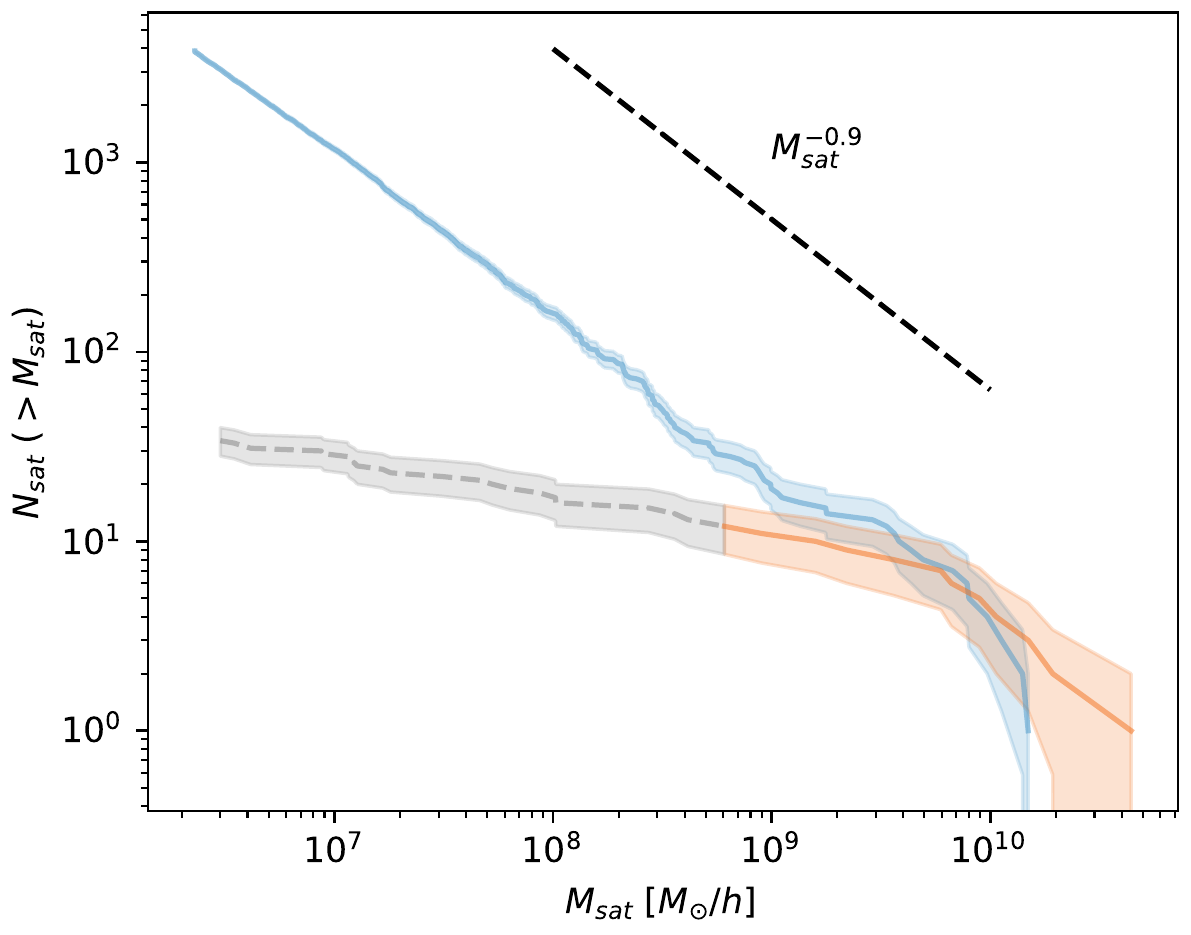}
\includegraphics[width=0.49\textwidth, trim={0cm 0cm 0.1cm 0cm}, clip] {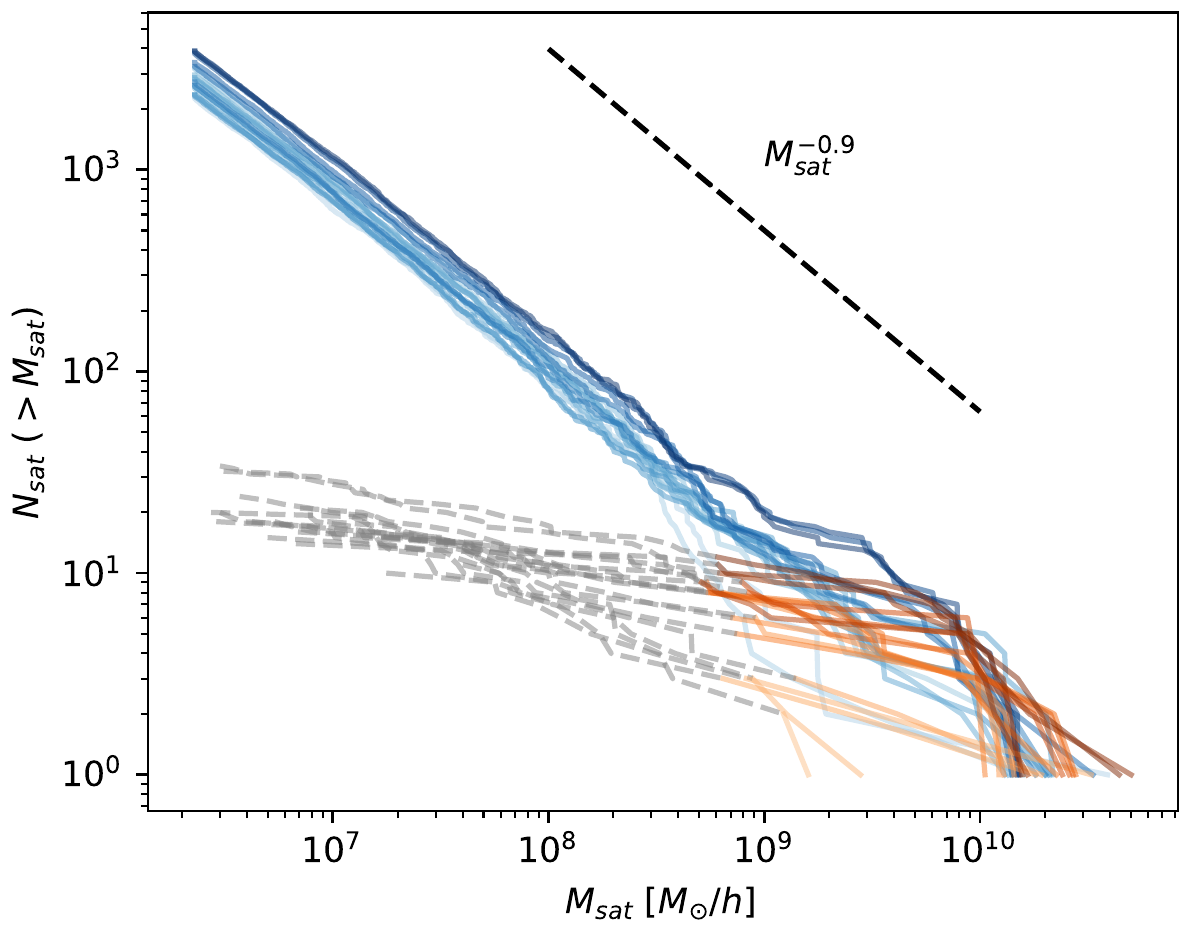}
\caption{Subhalo mass function of the CDM (blue) and FDM (orange) systems, at redshift $z=0$ (left) and at multiple redshifts (right). In the latter, different redshift from $z=4$ and $z=0$ are identified by a gradient in the color shade, from lightest to darkest. In the left panel we also show the Poissonian error (shaded region), not present in the right panel for readability purposes. The FDM subhalo mass function is represented by a grey dashed line for masses below the mass cut $M_{CUT}$.}
\label{fig:shmf}
\end{figure*}

The impact of such \textit{healing} process is even more noticeable in the evolution of the SubHalo Mass Function (SHMF). The SHMF at redshift $z=0$ is displayed in the left panel of Fig.~\ref{fig:shmf} for FDM (orange) and CDM (blue), respectively, with the shaded region representing the Poissonian error associated with the finite number of objects. To better visualise the evolution of the SHMF in time, in the right panel of Fig.~\ref{fig:shmf} the SHMFs are displayed at various redshift using a gradient in the color shades --~ranging from the lightest shade at $z=4$ to the darkest at redshift $z=0$~--. Poissonian error are omitted in this case to improve readability. The mass cut imposed in the FDM case is visualized by a change in color and linestyle (solid orange to dashed grey).

Comparing the evolution of the SHMFs, it is possible to see that the number of CDM \sats grows consistently across all mass bins maintaining the scaling $M_{sat}^{-0.9}$ \citep[one of the main results of the original Aquarius paper][]{Aquarius}. On the contrary, the number of \sats in FDM --~starting well below the CDM counterpart due to the power suppression in the initial conditions~-- evolves unevenly, with the higher mass end growing much faster than the lower end and effectively overlapping with the CDM SHMF at low redshift, due to the gravitational healing effect. 

\bigskip

To summarise, these results on the number and mass of \sats confirm that in FDM the number of \sats is reduced, in particular at the expenses of the smallest \sats. When considering the most massive \sats, the reduction in number and mass share is less pronounced with respect to CDM in the same mass range, altough still observable. Even though these massive \sats are the most likely to host an observable bright baryonic counterpart, the considerable departure of the HMFs is very likely to translate in an observable difference in the luminosity function \citep[as shown for other dark matter models as WDM in][]{MaccioFontanot10}.

\subsubsection{Distance and velocity distribution}
\label{sec:distancevelocity}

The different dynamics introduced in the FDM scenario has an impact not only on the number and mass of \sats, as we detailed in the previous subsection, but also on the overall properties of \sats in terms of position and velocity. 

\begin{figure*}
\includegraphics[width=0.70\textwidth%,trim={1cm 2.1cm 1cm 1cm},clip
]{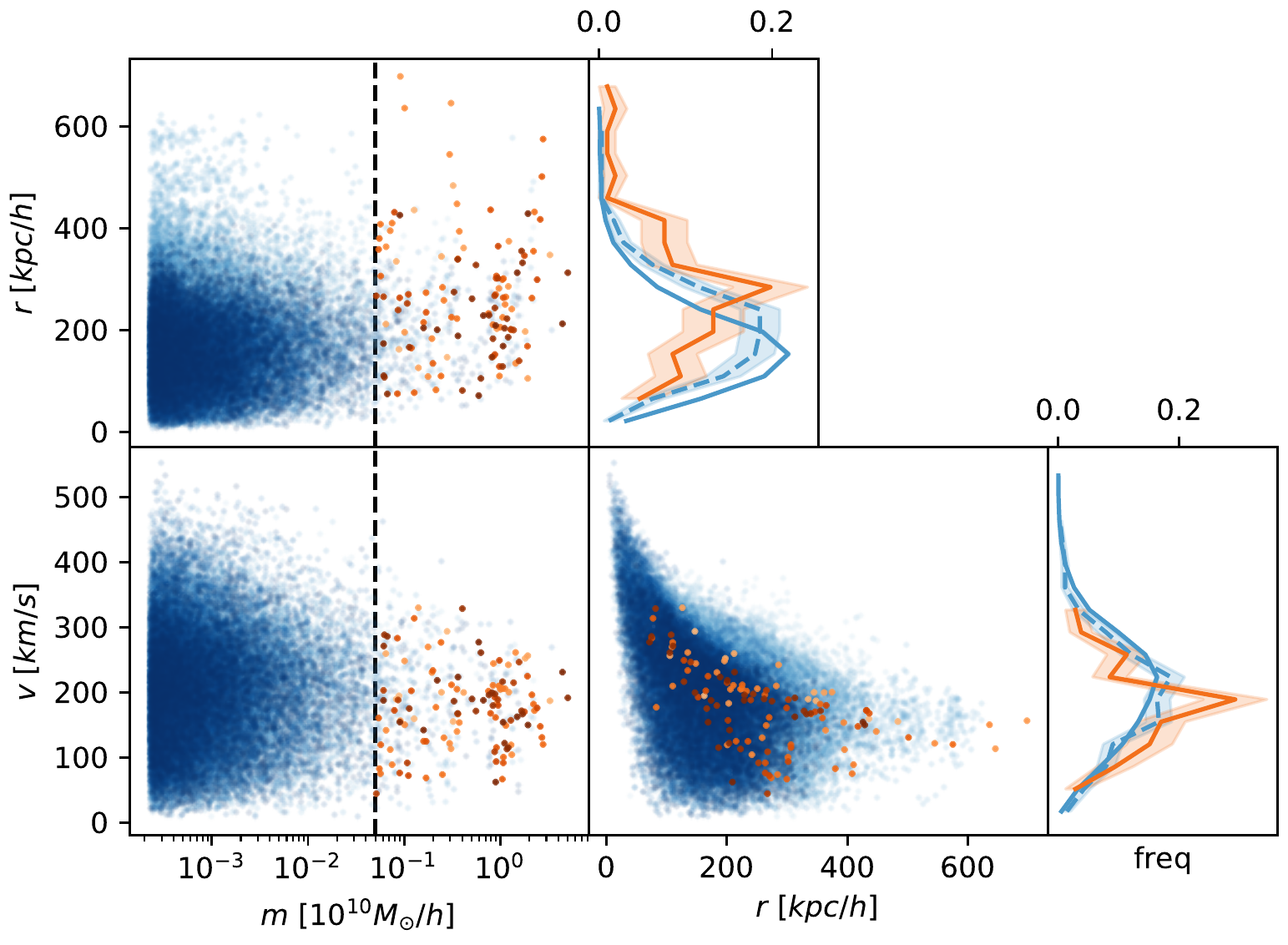}
\caption{Distributions of \sats properties in terms of distance, velocity and mass in the FDM (orange) and CDM (blue) simulations. The vertical dashed line in left panels represents the mass cut $M_{CUT}$. Integrated distributions of \sats in distance and velocity in the two simulations are featured on the right side each with its own shaded $1\sigma$ confidence region, where the distribution of CDM \sats with $M_{sat} > M_{CUT}$ is added for comparison (dashed blue). To avoid a very low count statistics in the FDM case, \sats in snapshots at redshift $z<1$ are here stacked together.}
\label{fig:rvm}
\end{figure*}

The pair-wise distributions of distance, velocity and mass are displayed in Fig.~\ref{fig:rvm} where points in the parameter spaces represents \sats found at redshift $z<1$. Integrated distributions in mass are collected in the right panels (the stacking on redshift is meant to avoid a low-count statistics in the FDM case). The FDM (orange) and CDM (blue) points at various redshifts are once again depicted using a gradient in the colour shading.

The vertical dashed line in the left panels represents the mass cut $M_{CUT}$ applied to the FDM and CDM-CUT samples. For comparison, the integrated distributions --~each depicted with its own shaded $1\sigma$ confidence region~-- are plotted alongside the integrated distribution of the CDM-CUT subsample that satisfy $M_{sat} > M_{CUT}$ (dashed blue line).

In the FDM scenario, \sats can be found at larger distances with respect to CDM-CUT, with a statistical shift of $\sim50$~kpc/h separating the peak of the two distributions obtained in the same mass range (note also that the peak of the CDM-CUT distribution itself is $\sim50$~kpc/h shifted farther away from the center with respect to the total CDM one). Other than the peak itself, the difference in the distributions seems to be more prominent in the tails: FDM \sats are statistically found at larger distances than CDM and are scarse in the vicinity of the center of the system. The velocity distributions are statistically consistent with each other, with a small preference for low velocity in FDM with respect to the CDM case, although not dramatically significant.

Ultimately, the combined analysis of these distributions confirms that \sats in FDM are statistically found at a larger distance from the centre and do not reach the high velocities observed in CDM. This is due to the lack of low-mass \sats that in CDM case contribute to the high velocity and to small distance tails of the two distributions. Low-mass \sats in FDM are in general less numerous than the ones in CDM as small over-densities are not present in the initial conditions. However, the few low-mass \sats that are indeed able to form in FDM are found statistically farther away from the \main and have a lower velocity. 
This is a strong indication that the properties of \sats are also affected by dynamic processes in the evolution of the system, most possibly through an enhanced stripping effect  exerted by the \main on small and close \sats in the FDM cosmology. Full hydro simulations will be needed to probe the effects of the different dynamical evolution of the stellar content of the satellites, since dark matter and stars react differently to stellar stripping \citep{Penarrubia2008,Maccio2021_galaxywithoutdarkmatter}.

\subsubsection{Properties of the \main}
\label{sec:mainsubhalo}

Let us now focus on the properties of the \main. Given its high mass and large size, its global properties are less affected by the additional physics of FDM --~that has only an impact on the inner region~-- with respect to \sats. In fact, the total mass of the \main in the FDM and CDM scenarios are comparable at all redshifts within a $\pm20\%$ range, with the largest deviations coming from different timings in the major merger events. Nevertheless, a deviation in the radial profiles induced by the typical repulsive FDM interaction is indeed expected.

\begin{figure}
\includegraphics[width=\columnwidth%,trim={1cm 2.1cm 1cm 1cm},clip
]{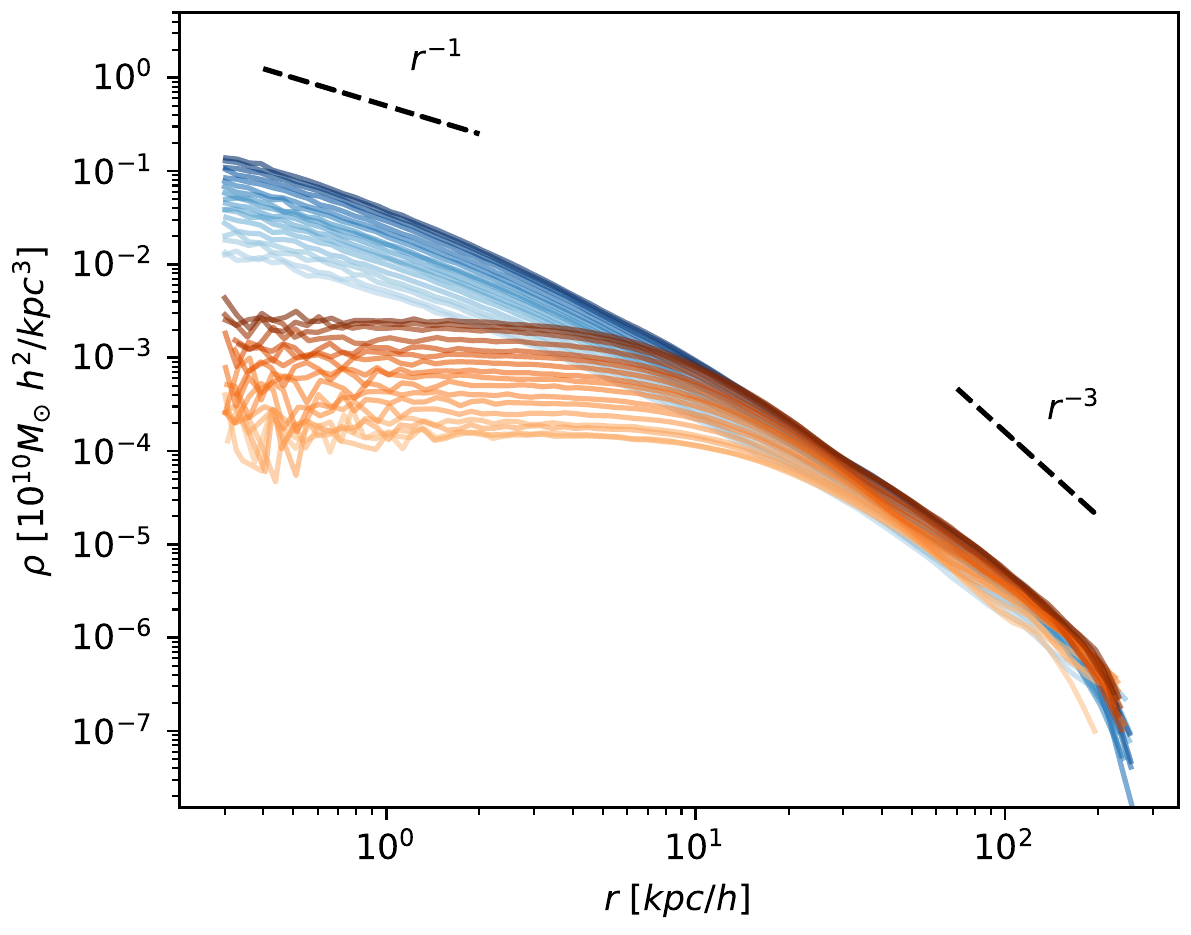}
\caption{Radial dark matter density profiles of the \main in FDM (orange) and CDM (blue). Profiles at different redshifts are displayed with a gradient in shading, as in Fig.~\ref{fig:shmf}.}
\label{fig:dprof}
\end{figure}

In Fig.~\ref{fig:dprof}, the radial density profile of the \main is displayed both for the FDM (orange) and CDM models (blue). For the two scenarios, profiles at multiple redshifts are displayed with a shade gradient --~i.e. going from lighter to darker shades from $z=2.5$ to $z=0$~-- to visually represent the evolution of the profile in time.

As clearly visible, while the CDM profile is consistent with a NFW profile, the FDM one features the typical combination of a core in the centre and a NFW tail. This comes as no surprise, as it is known that the stable configuration of FDM haloes is indeed characterised by a core sustained by the repulsive effect of the QP, as described in Sec.~\ref{sec:fdm_sr}. 

\begin{figure}
\includegraphics[width=\columnwidth%,trim={1cm 2.1cm 1cm 1cm},clip
]{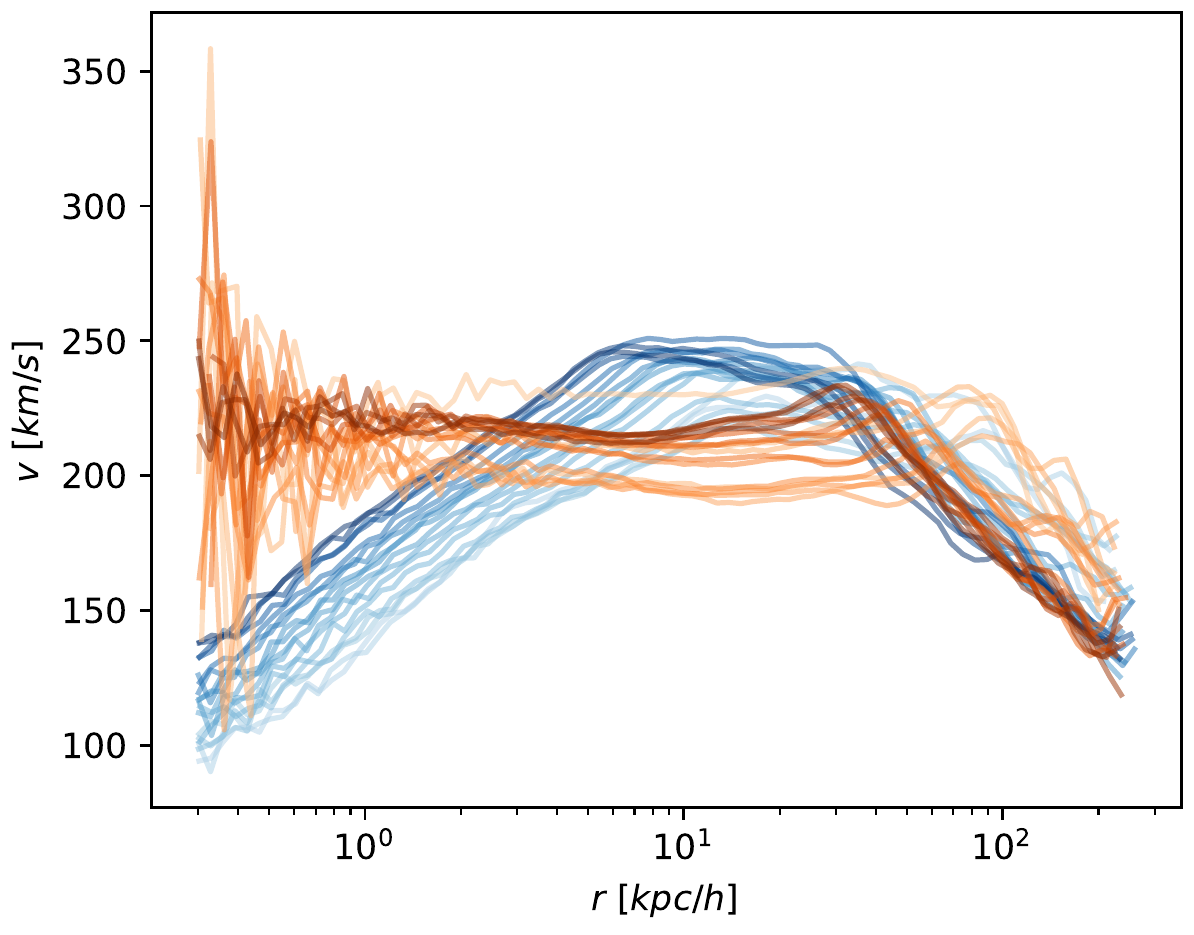}
\caption{Radial dark matter velocity profiles of the \main in FDM (orange) and CDM (blue). Profiles at different redshifts are displayed with a gradient in shading, as in Fig.~\ref{fig:shmf}.}
\label{fig:vprof}
\end{figure}

In this cored configuration, the phase of the wave-function --~related to the fluid velocity in the Madelung formulation~-- is position independent, thus a core in the density distribution is paired with a \textit{plateau} in the velocity distribution \citep[]{Hui16}. Heuristically, as the FDM core is defined as the solution for which the gravitational self-attraction is compensated by an opposite force at all points, no velocity gradient is needed for the core self-sustainment, contrary to the CDM case. To highlight this peculiar effect, the velocity radial profiles are shown in Fig.~\ref{fig:vprof} for the \main in the CDM (blue) and FDM (orange) scenarios, with the shade gradient to represent different redshift as in Fig.~\ref{fig:dprof}.

Regarding the CDM model, the velocity profile is consistent with an NFW density profile. In fact, in CDM we can use the following qualitative estimate: assuming virialisation, the velocity profile $v(r)$ given by a power-law density distribution $\rho(r)\sim r^{-n}$ is related to the mass $M(r)$ enclosed within $r$ as $v(r)\sim \sqrt{M(r)/r}$, thus $v(r)\sim r^{-n/2+1}$. Since the NFW density profile is characterised by a double power-law behaviour with inner $\rho(r)\sim r^{-1}$ and outer $\rho(r)\sim r^{-3}$ scalings, the velocity profile observed $v(r)\sim r^{+1/2}$ and $v(r)\sim r^{-1/2}$ are indeed consistent with predictions.

In FDM, instead, it is not possible to invoke the same simple estimate as the scaling $v(r)\sim \sqrt{M(r)/r}$ is obtained by assuming equilibrium between the kinetic and gravitational energy only, while the quantum energy typical of FDM should be included in this case. In the outskirts of the \main of the NFW-like tail, the velocity profiles of FDM and CDM are consistent with each other as the quantum energy contribution is negligible in the integration --~being dominant only in the innermost region~--, while inside the core the different dynamics translates in a velocity \textit{plateau}, as previously discussed.

While the inner cored feature in the dark matter density profile can be reproduced dynamically by many dark matter scenarios alternative to CDM, this very interesting feature in the velocity profile clearly sets FDM apart and could be regarded as a typical feature. Indeed, even though it may be hard to observe, this is a clean feature to discriminate CDM and other dark matter models --~as e.g. thermal WDM~-- from FDM. Of course, given that this work is restricted to a dark-matter-only scenario, this peculiar feature might be potentially observable in astrophysical systems where the baryonic content is negligible, as dwarf galaxies or smaller systems.

\subsection{Specific FDM core properties}

In the previous section, we verified that the \main hosts a FDM core associated with a velocity plateau, identifying the solution of the spherical mEP system of Eq.~\ref{eq:mEP}. In this section, we now tackle the specific properties of the core in more detail, including the \sats in this analysis. 

Before presenting the results, let us first recall our expectations about the solitonic state of a FDM core. In a purely ideal and isolated case, a spherically symmetric core solution in an excited state should eventually relax and reach the solitonic state --~i.e. the ground state of the system~--, exhibiting a purely quantum transition between discreet excitation levels associated with sudden energetic emissions. Of course, the discreet nature of this quantum process is less dramatic when considering a not-idealised and rather chaotic physical system such as the formation of dark matter structures, with a large number of degrees of freedom --~e.g. intrinsic and collective motion, morphological asymmetry, etc.~--, so that excited systems are able to slowly dissipate energy and their relaxation process to reach the ground-state is rather smooth. 

It is thus reasonable to imagine that the inner cored structure forms in the early stages of the overall system collapse as a result of the competing gravitational and quantum potentials, yet not necessarily in its energetic ground-state. In time, with the system relaxation and assuming no external energy input, the core may eventually dissipate its excess energy and gradually reach its solitonic state.

\begin{figure}
%\frame{
\includegraphics[width=\columnwidth,trim={0.5cm 0cm 1.8cm 1.5cm},clip]
{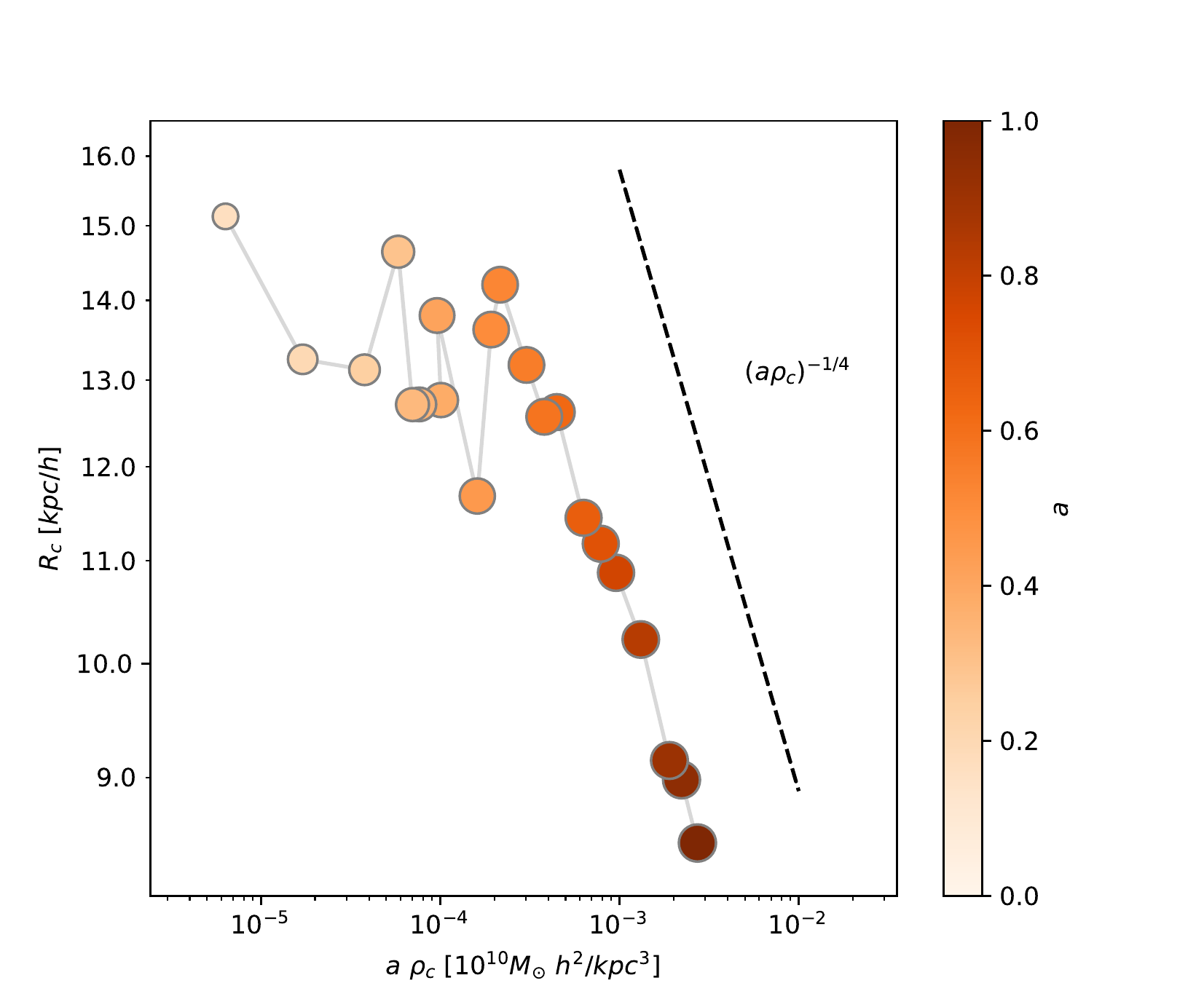}
\caption{Evolution of the core properties in the $(\rhoc, \Rc)$ parameter space. The scaling $\Rc \sim (a \rhoc)^{-1/4}$ is here given as a reference for the slope with no particular normalisation.}
\label{fig:core}
\end{figure}

Having this in mind, let us focus on the trajectory of the core of \main in the $(\rho, \Rc)$ parameter space, represented in Fig~\ref{fig:core}. Here, only the direct progenitor line is considered, meaning that no secondary progenitor branchings are depicted. The position of the core properties in this parameter space is once again displayed for several redshifts from $z=4$ to $z=0$ using a shade gradient from lightest to darkest.

It is possible to see that evolution of the \main core properties in time is characterised by an initial \textit{stalling} phase (roughly from $z=4$ to $z=1$ in this case) when the density increase is accompanied by only a marginal decrease in radius. After that, the radius decrease more rapidly and the typical scaling $\Rc \sim (a  \rhoc)^{-1/4}$ --~presented in the plot as a dashed line with no particular normalisation~-- is reached and maintained until the end of the simulation.

As another important result of this work, we observe that the process for the core to reach a stable scaling does not seem to be necessarily instantaneous. Of course, the following question naturally arises: are there processes that are able to delay the onset and the eventual stabilisation of the expected core scaling? 

To tackle this question, let us broaden the sample under consideration and include the \sats in this analysis. All $34$ satellites in our simulations feature a FDM core and serve as valid solutions of the mEP system. Using the conservative cuts $M_{CUT}= 5 \times 10^8 \dimM$ and $s_{CUT}=0.16$ to account for numerical fragmentation, we restrict our analysis on $12$ out of the $34$ \sats belonging to the system at redshift $z=0$, noting that the remaining $22$ represent $<3\%$ of the total mass found in \sats at that redshift. In the following, we are going to study the \main and \sats cores as part of a single sample, not taking in consideration the great difference in age and environment within which these objects have evolved. We acknowledge that this might be a very important factor, but this is something that only a broader study encompassing a large sample of complex systems (with data on \sats) can address, thus falling beyond the scope this work.

\begin{figure*}
\includegraphics[width=\textwidth, trim={1.6cm 0.8cm 3.4cm 1.1cm}, clip] {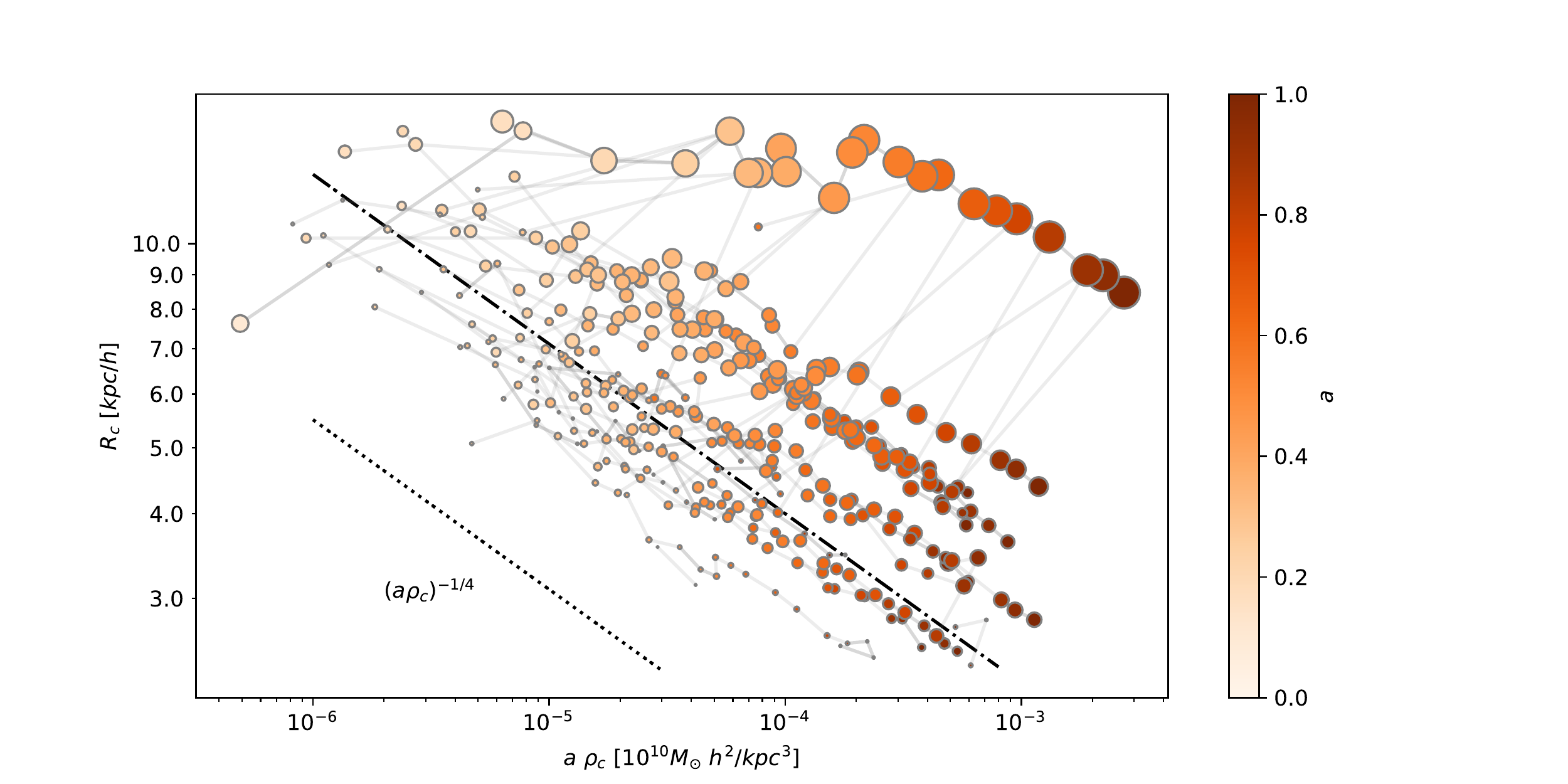}
\includegraphics[width=\textwidth, trim={1.6cm 0.0cm 3.4cm 1.1cm}, clip] {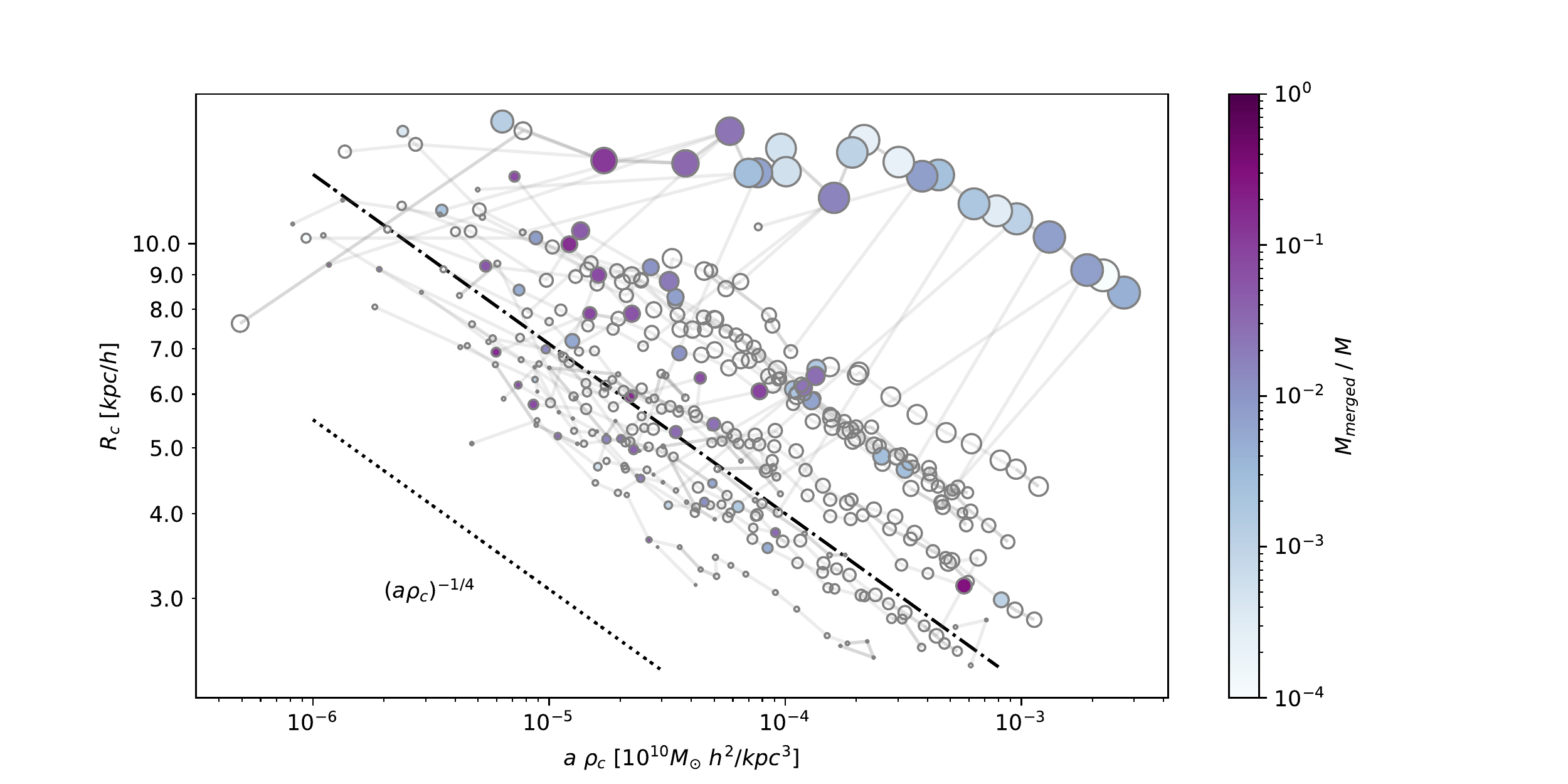}
\caption{Evolution of the core properties in the $(\rhoc, \Rc)$ parameter space space. Point size is related to mass, while the color represent a different observable for each panel; these are the scale factor $a$ (top panel) and the fraction of mass accreted via merger $M_\text{merged}/M$ from the previous redshift (bottom panel). The black lines represent the typical $\Rc \sim (a  \rhoc)^{-1/4}$ scaling with \citet{Schive14} (dotted) and \citet{Nori20} (dotted dashed) normalisation.}
\label{fig:core_multi}
\end{figure*}

In Fig.~\ref{fig:core_multi}, the position in the $(\rhoc, \Rc)$ parameter space of the \main and the selected \sats are displayed, including the secondary progenitor branches recovered from the merger tree analysis (progenitors and descendants in the figure are linked by faint grey lines). As reference, the $\Rc \sim (a  \rhoc)^{-1/4}$ scalings with the normalisation of \citet{Schive14} (dotted line) and \citet[][]{Nori20}\footnote{Specifically $\kappa_{1/4}$ from Table 4 \citep[][]{Nori20}.} (dot-dashed line) are plotted.

The point size is related to the halo mass, while color is representative of a different observable for each panel: these represent the scale factor $a$ in the top panel and the fraction of mass accreted via merger from the previous snapshot $M_\text{merged} / M$ in the bottom one, with $M_\text{merged}$ as defined in Sec.~\ref{sec:subfind}. 

Regarding the distribution of the core properties in the parameter space and its relation with redshift (first panel of Fig.\ref{fig:core_multi}), it is possible to notice that trajectories are characterised by four overall features:
\begin{itemize}
    \item all cores eventually reach the expected $\Rc \sim (a \rhoc)^{-1/4}$ scaling in terms of slope;
    \item the normalisation factor of the final scaling relation is not the same for every core, seemingly showing a positive correlation with the system mass and/or age;
    \item cores of similar mass and forming at similar redshift seem to share similar trajectories; 
    \item in few cases, trajectories exhibit sudden and abrupt deviations, especially at the nodes of merger tree.
\end{itemize}

The results in Fig.~\ref{fig:core_multi} suggest that all cores end up following the $\Rc \sim (a \rhoc)^{-1/4}$ scaling at lower redshifts. In terms of normalisation, more massive \sats tend to follow a scaling with a higher normalisation than the less massive ones, with a factor $\sim5$ of difference over almost four orders of magnitude in mass. This higher normalisation factor is also coupled with a more pronounced initial phase of marginal radius decrease (as noted in the \main). More massive \sats that formed at a higher redshift thus take statistically more time to reach their final scaling with respect to low mass (and low redshift) ones. 

Interestingly, the satellites in the same mass range of \citet{Nori20} are consistent with the scaling found in that work, proving the consistency of the results of different systems and different resolutions. In \citet{Nori20}, this scaling was studied considering a collection of zoom-in systems with different masses across two orders of magnitude, finding that the overall scaling relation was generally tilted and influenced by the state of relaxation of the systems. Given the results obtained in this work, we can conclude that the tilt in the overall scaling relation observed in \citet{Nori20} originated from calculating a single slope using systems that reached the same scaling relation but at different times and with different normalisations (it is possible indeed to recognise \textit{a posteriori} this effect in the left panels of Fig.6 in \citet{Nori20}). 

The sudden deviation that some trajectories exhibit are preferentially found at merger events --~highlighted in the bottom panel~--. Indeed, it would be reasonable to claim that merger events might play an important role in delaying the realisation of the scaling and even temporarily disrupting it. The amount of mass accreted via merger also appears to be more relevant in the initial \textit{stalling} phase, supporting a potential role in this sense. 

However, it is important to stress that the hierarchical formation of structures --~that holds also in FDM cosmologies~-- comes with an intrinsic positive correlation between the mass of objects, their age and the number/importance of merger events they experienced, so it impossible at this stage to conclude which one of these properties has a direct causal effect on the normalisation of the FDM scaling. Moreover, even if this correlation is physically reasonable and seems to be consistently present in simulations of different systems and resolutions, it is impossible to exclude --~as well as to confirm~-- a possible numerical origin, mainly related to the SPH nature of the algorithm. This in fact could be only tested in a direct comparison between simulations of the same system with a particle based and a grid based approach, which still is somehow difficult to design due to the different numerical limits on scale and resolution that the two methods have. 

\section{Conclusions}
\label{sec:conclusions}

In this work, we presented a high-resolution simulation of the formation and evolution of a Milky-way-like halo extracted from the Aquarius project in the framework of Fuzzy Dark Matter (FDM) cosmologies. With the use of the N-body code \AG, we were able to simulate such complex system, with a massive central object and a considerable population of satellites, without neglecting the typical quantum interaction of FDM.

We first detailed the global properties of \sats, presenting --~for the first time in the literature~-- a self-consistent Subhalo Mass Function in a FDM system. In the comparison with its CDM counterpart, the overall number of \sats is greatly reduced, especially for the least massive ones. The total mass found in \sats is reduced as well. In terms of distance and velocity, \sats in FDM are statistically found at larger distances and have lower speed.

With the relatively low FDM mass $m_{\chi}=2.5h \times 10^{-22}\ {\rm eV}/c^2$ adopted in this work --~chosen to highlight the effect of FDM dynamics on galaxy formation processes~--, the number of dark matter \sats in FDM is approximately the same of the luminous satellites of a Milky-way-like galaxy \citep{McConnachie2012}. Assuming that such FDM model is valid would require almost all of the dark \sats to have a luminous component, which the current knowledge of structure formation disfavours. Of course, a larger value of the FDM mass would enhance the number of dark matter \sats, thus representing a viable model in terms of number of luminous counterparts.

The \main exhibits the typical cored structure of FDM in the radial density profile. Its dark matter radial velocity profile is characterised by a flattening in the core region: this feature, that to our knowledge is observed in this work for the first time, is indeed a specific feature of FDM that sets the model apart from other dark matter models such as Warm Dark Matter, although being hardly observable at the moment. 

Studying the evolution of the radius and density of this core, we observe that the theoretical scaling relation $\Rc \sim (a  \rhoc)^{-1/4}$ is asymptotically approached, after an initial stalling phase where the radius only marginally decreases.

Finally, the evolution of the core properties of the \main was framed in a broader context by studying the combination of the core in the \main as well as the cores of \sats. The properties of all cores satisfy the scaling relations, although the specific normalisation was unexpectedly found to mildly correlate with the system mass --~or equivalently with age or merger events, as these properties are all related to each other~--.

This work shows that the properties of cores that theoretically follow rather simple FDM scaling relations exhibit a more complex behaviour, characterised in some cases by a delayed establishment of the scaling relation (and not immediate from formation onward), which is also not necessarily universal in terms of normalisation when systems with many orders of magnitude of difference in mass and are considered. The interplay between subsystems and the peculiar history of each subsystem can indeed play a role in such differentiation. These effects seem to be more pronounced for the most massive systems in our analysis, thus suggesting that the indiscriminate use of the scaling relations to confirm or rule out FDM models in the comparison with observations might be incorrect when the extrapolation extends indefinitely over various orders of magnitude in mass.

\section*{Acknowledgements}

The authors thank Prof. Adrian Jenkins, who provided the original Aquarius CDM simulation and helped the authors in the realisation of the initial conditions for FDM. This material is based upon work supported by Tamkeen under the New York University Abu Dhabi Research Institute grant CAP$^3$. MB acknowledges support by the project ``Combining Cosmic Microwave Background and Large Scale Structure data: an Integrated Approach for Addressing Fundamental Questions in Cosmology'', funded by the MIUR Progetti di Ricerca di Rilevante Interesse Nazionale (PRIN) Bando 2017 - grant 2017YJYZAH. The authors gratefully acknowledge the High Performance Computing resources at New York University Abu Dhabi and at the parallel computing cluster of the Open Physics Hub (\url{https://site.unibo.it/openphysicshub/en}) at the Physics and Astronomy Department in Bologna. 

%The authors acknowledge support from the Italian Ministry for Education, University and Research (MIUR) through the SIR individual grant SIMCODE, project number {RBSI14P4IH}. The simulations presented in this work were performed on the Marconi supercomputer at CINECA. The authors acknowledge the ISCRA initiative for support and resources (two class C allocations, project numbers: HP10CPVK72 and HP10C0JGGH).% This research was also supported by the Munich Institute for Astro- and Particle Physics (MIAPP) which is funded by the Deutsche Forschungsgemeinschaft (DFG, German Research Foundation) under Germany's Excellence Strategy - EXC-2094 - 390783311. 

\section*{Data Availability}
The data underlying this article will be shared on reasonable request to the corresponding author.

%%%%%%%%%%%%%%%%%%%%%%%%%%%%%%%%%%%%%%%%%%%%%%%%%%



\bibliographystyle{mnras}
\bibliography{BIB,baldi_bibliography} % if your bibtex file is called example.bib

%%%%%%%%%%%%%%%%%%%%%%%%%%%%%%%%%%%%%%%%%%%%%%%%%%

%%%%%%%%%%%%%%%%% APPENDICES %%%%%%%%%%%%%%%%%%%%%

%\appendix

%\section{Soliton fit}

%%%%%%%%%%%%%%%%%%%%%%%%%%%%%%%%%%%%%%%%%%%%%%%%%%

% Don't change these lines
%\bsp	% typesetting comment
\label{lastpage}
\end{document}